\documentclass[namedreferences]{solarphysics} 

\usepackage[hyperref,optionalrh]{spr-sola-addons} 

\usepackage[a4paper,total={210mm, 297mm},top=10mm, bottom=45mm,left=35mm, right=35mm]{geometry}

\usepackage{graphicx}        
\usepackage{epstopdf}
\usepackage{xcolor}           
\usepackage{breakurl}        
\usepackage[british]{babel} 
\usepackage[T1]{fontenc}
\usepackage[utf8]{inputenc}

\usepackage[leftcaption]{sidecap} 
\sidecaptionvpos{figure}{t} 
\usepackage{rotating}
\usepackage{hyperref}
\hypersetup{colorlinks,linkcolor={blue},citecolor={blue},urlcolor={red}}
\usepackage{multirow}

\makeatletter

\usepackage{float}

\usepackage{tikz}
\usetikzlibrary{arrows.meta,bending}
\usetikzlibrary{shapes.misc}
\tikzset{cross/.style={cross out, draw=black, minimum size=2*(#1-\pgflinewidth), inner sep=0pt, outer sep=0pt},
	cross/.default={3mm}}

\newcommand{\arcsec}{\mbox{$^{\prime\prime}$}}%
\def\dif{\: {\rm d}}                       

\def\Halpha{\mbox{H\hspace{0.1ex}$_\alpha~$}}
\def\deg{\hbox{$^\circ$}}

\newcommand{\eg}{{\it e.g.}}

\newcommand{\etc}{{\it etc.}}
\newcommand{\ie}{{\it i.e.}}

\newcommand{\aap}{    {\it Astron. Astrophys.}}

\newcommand{\apj}{    {\it Astrophys. J.}}                   

\newcommand{\jfm}{    {\it J. Fluid Mech.}}

\newcommand{\mnras}{  {\itshape Mon. Not. Roy. Astron. Soc.}}

\begin{document}
	\begin{article}
		\begin{opening}
			
			\title{Magnetic and Velocity Field Topology in Active Regions of Descending Phase of the Solar Cycle 23}
			
			\author[addressref={aff1},corref,email={ramajor@nitc.ac.in}]{\inits{R.~A.}\fnm{R.~A.}~\lnm{Maurya}}
			\address[id=aff1]{Department of Physics, National Institute of Technology Calicut, Calicut-673601, India.}
		
			\author[addressref={aff3},corref,email={ambastha@prl.res.in}]{\inits{A.}\fnm{A.}~\lnm{Ambastha}}
			\address[id=aff3]{Udaipur Solar Observatory, Physical Research Laboratory, Udaipur-313001, India.}
			
			\runningauthor{Maurya and Ambastha}
			\runningtitle{Topology of Solar Active Regions}
			
			\makeatother
			
			\begin{abstract}
				
				We analyse the topology of photospheric magnetic fields and sub-photospheric flows of several active regions (ARs) that are observed during the peak to descending phase of the solar cycle 23. Our analysis shows clear evidence of hemispheric preferences in all the topological parameters such as the magnetic, current and kinetic helicities, and the `curl-divergence'. We found that 68\%(67\%) ARs in the northern (southern) hemisphere with negative (positive) magnetic helicities. Same hemispheric preference sign is found for the current helicities in 68\%(68\%) ARs. The hemispheric preferences are found to exist statistically for all the time except in few ARs observed during the peak and the end phases of the solar cycle. This means that magnetic fields are dominantly left(right)-helical in scales smaller than individual ARs of northern(southern) hemisphere. We found that magnetic and current helicities parameters show equator-ward propagation similar to the sunspot cycle. The kinetic helicity showed similar hemispheric trend to that of magnetic and current helicity parameters. There are 65\%(56\%) ARs with negative (positive) kinetic helicity as well as divergence-curl, at the depth of 2.4\,Mm, in the northern (southern) hemisphere. The hemispheric distribution of the kinetic helicity becomes more evident at larger depths, \eg, 69\%(67\%) at the depth of 12.6\,Mm. Similar hemispheric trend of kinetic helicity to that of the current helicity supports the mean field dynamo model. We also found that the hemispheric distribution of all the parameters increases with the field strength of ARs. The topology of photospheric magnetic fields and near surface sub-photospheric flow fields did not show good association but the correlation between them enhances with depths which could be indicating more aligned flows at deeper layers of ARs.     
				
			\end{abstract}
			
		\end{opening}
		
		\section{Introduction}
		\label{sec:intro}
		
		Solar active regions are three-dimensional magnetic structures extending from the interior below the photosphere to the coronal heights. They consist of magnetic flux tubes that emerge from the convection zone and extend into the solar atmosphere \citep[][and references therein]{Hood2011}. Solar interior near the photosphere is an excellent conductor where frozen-in-field condition may satisfy \citep{Priest2014}. Over a period of time, near the photosphere, plasma flows may evolve with magnetic field lines, and may swirl around them.  A flux tube emerging from the convection region is prevented from being destroyed by the hydrodynamics vortex behind it when twisting effects are present \citep{Fan1998}. Therefore, we can expect topological association between the sub-photospheric flows and photospheric magnetic fields.

		Mathematically, the topology of a vector field $\mathbf{X}$ is represented by the parameter ``helicity''. It is defined by the volume integral of the scalar product of the vector field and its curl,  $\mathbf{Y}=\nabla\times\mathbf{X}$, 
		
		\begin{equation}
		H=\int_{V}\mathbf{X}\cdot\mathbf{Y}\dif V.
		\label{eq:hel-def}
		\end{equation}
		
		\noindent The helicity is a fundamental topological quantity in laboratory as well as in astrophysical plasma. In general, helicity (Equation~\ref{eq:hel-def}) describes what kind of handedness is preferable for the field $\mathbf{X}$, left or right hand screw. If $\nabla\times\mathbf{X}$ is clockwise when viewed from ahead of the body, then helicity is positive; and if counter-clockwise, it is negative. The helicity defined in Equation~\ref{eq:hel-def} is a global parameter over the entire volume $V$. 
		
		The magnetic helicity of a divergence-free field $ \mathbf{B} $ within a volume $ V $, bounded by a surface $ S $, is defined by \citep{Elsasser1956},

		\begin{equation}
		H_{\rm m} = \int_{\rm V}\mathbf{A}\cdot\mathbf{B}\dif V
		\label{eq:mag-hel-def}
		\end{equation}

		\noindent where, the vector potential $\mathbf{ A} $ satisfies, $\mathbf{B}=\nabla\times\mathbf{A}$. However, Equation~\ref{eq:mag-hel-def} is gauge invariant (independent of  $ \phi $ where $ \mathbf{A^\prime} = \mathbf{A} + \nabla \phi $) when the magnetic field is fully contained inside the volume $ V $. The helicity $H_{{\rm m}}$ is a measure of linking, knotting and twisting of field lines \citep{Moffatt1969, Berger1984a}.  Note that there is a subtle mathematical difference between the general definition of the helicity $ H $ (Equation~\ref{eq:hel-def}) and the magnetic helicity $H_{{\rm m}}$ (Equation~\ref{eq:mag-hel-def}).  In the definition of the helicity $ H $ of the field $ \mathbf{ X} $,  we integrate the scalar product of the field $ \mathbf{ X} $ with its curl, while for the helicity $H_{{\rm m}}$ of magnetic field $ \mathbf{ B} $, we integrate the scalar product of vector potential  $ \mathbf{ A}$ of magnetic field $ \mathbf{ B} $ (and NOT $ \mathbf{ B} $ itself)  with its curl. This makes difficult to compute the magnetic helicity, as the vector potential $ \mathbf{ A} $ is gauge invariant and the full 3D magnetic fields are not observable. Therefore, force-free parameter $ \alpha_f $  given in the equation,
		
		\begin{equation}
		\nabla\times \mathbf{ B} = \alpha_f \mathbf{ B},
		\label{eq:alpha}
		\end{equation}
		
		\noindent is taken as a proxy for the magnetic helicity by several researchers \citep[][and references therein]{Pevtsov2008}. The topology of flow fields can be given by the kinetic helicity \citep{Moffatt1992, Moffatt1969}. It is defined as the volume integral of the scalar product of velocity ($\mathbf{u}$) and vorticity {\boldmath$\omega$} , 
		
		\begin{equation}
		H_{{\rm k}}=\int_{{\rm V}}\mathbf{u}\cdot\boldmath{\omega}\dif V
		\label{eq:kin-hel-def}
		\end{equation}
		
		\noindent where, vorticity, {\boldmath$\omega=\nabla\times\mathbf{u}$}, is another important topological parameter to describe the flow which represents circulation per unit area or twist of fluid fields. It is a pseudo-vector (or axial vector) that transforms like a vector under a proper rotation, but in three dimensions again an additional sign flip occurs under an improper rotation such as a reflection.  Note that $H_{{\rm k}}$ retains the same mathematical form for the flow field $ \mathbf{ u} $ as of $ H $ for the vector field $ \mathbf{ X} $. It represents extent to which corkscrew-like motion occurs. If a parcel of fluid is moving and rotating about an axis parallel to the direction of motion, it will be positive. In the absence of complete velocity field observations, we compute the kinetic helicity density, 
		
		\begin{equation}
		h_{k}=\mathbf{u}\cdot\omega=\mathbf{u}\cdot\nabla\times\mathbf{u},
		\label{eq:kin-hel-den}
		\end{equation}
		
		\noindent as a proxy for the swirls in the sub-photospheric flows. The kinetic helicity density, $ h_k $, is a scalar quantity and it represents the component of vorticity along the direction of flow. In magneto-hydrodynamics, the kinetic helicity of flow field has similar mathematical form as that of the ``current helicity'',
		
		\begin{equation}
		H_{{\rm c}}=\int\mathbf{J}\cdot\mathbf{B}\dif V
		\label{eq:cur-hel-def}
		\end{equation}
		
		\noindent where, $\mathbf{J}$ is current density which is related with magnetic field by formula, $\mathbf{J}=\frac{1}{\mu_{0}}\nabla\times\mathbf{B}$, and $\mu_{0}=4\pi\times10^{-7}\,{H m^{-1}}$ is the permeability of free space. The current helicity quantifies the location of twist and sheared non-potential structures in a magnetic field of ARs. It defines the degree of twist and linkage of electric currents.  Since full 3D magnetic fields are not observed, we consider the current helicity density as follow,
		
		\begin{equation}
		h_{\rm c} = (\nabla\times\mathbf{ B})\cdot \mathbf{ B}.
		\label{eq:cur-hel-den}
		\end{equation}
		
		It is evident from  Equation~\ref{eq:mag-hel-def} and~\ref{eq:cur-hel-def} that only current helicity density can be estimated from the vector magnetograms under the assumption of non-linear force-free field, and the calculation of magnetic helicity density can only be made under the more restrictive assumption of constant $ \alpha_f $ of force-free field. However, the force-free parameter $ \alpha_f $ has the same sign as that of magnetic field under the assumption of linear force field \citep{Hagyard1999a}. They are related by the formula, $ h_{\rm c} = \alpha_f B_z^2 $ \citep{Seehafer1990}, which implies that for a given AR, the current helicity has same sign as that of the magnetic helicity. However, in a recent study, \citet{Russell2019} conjectured that current and magnetic helicities having the same sign is not true in general.
		
		In the solar chromosphere, magnetic structures such as \Halpha vortices, filament chirality, \etc, manifest the existence of helicity in ARs. The topology in all the above features demonstrate a consistent pattern relating a large scale ordering of solar magnetic fields. From a study of chromospheric \Halpha images of 51 sunspots from 1901 to 1944, \citet{Hale1927} first reported the preference of hemispheric trend in the \Halpha vortices. Later, \citet{Richardson1941} included some more sunspots and reported $71\%$ $(66\%)$ with left-handed (right-handed) vortices in the northern (southern) hemisphere.  The hemispheric sign rule (HSR) has also been found in the chirality of chromospheric filaments \citep[\eg,][]{Martin1994,Pevtsov2003,Lim2009}.
		
		Vector magnetic field observations of solar ARs have revealed that on  average solar ARs have a small but statistically significant mean twist and they show hemispheric biases. The magnetic helicity \citep{Pevtsov1995,Longcope1998,Bao2000, Labonte2007, Yang2009, Zhang2013b, Liu2014} and the current helicity \citep[][among others]{Seehafer1990,Bao1998,Abramenko1996, Pevtsov2001, Hagino2005, Gosain2013} of ARs have been found to follow HSR. However, opposite HSR in the global force free parameter $\alpha_f$ has been reported earlier  during the beginning \citep{Bao2000} and decay phases \citep{Tiwari2009a, Pipin2019} of solar cycles. From an analysis of several ARs,  \citet{Zhang2006} pointed out that the ARs with strong magnetic fields show opposite hemispheric trend. \citet{Pevtsov2008} compared data from four different instruments and concluded that HSR sign change in some phases of the solar cycle is not supported at high level of significance. \citet{Choudhuri2004} have numerically shown the hemispheric preference of magnetic helicity parameter $ \alpha_f $. They also noticed opposite HSR at the beginning of solar cycle. However, there is also some debate about this rule.
		
		Twist in a magnetic flux tube can be formed due to: (1) Coriolis force acting on rising flux tube \citep{Holder2004} and $ \Sigma $-effect \citep{Longcope1998}. The Coriolis force tilts one leg of the tube towards equator and the other away. This deformation generates helical distortion in the tube's central axis. (2) Flux tube buffeted by the turbulence of convection outside the tube during its rise through the convection zone. The kinetic helicity of the turbulent flows in convection zone twists the rising flux tube. (3) Background poloidal field \citep{Choudhuri2003, Choudhuri2004}. The magnetic helicity is expected to change its sign too because flows drags field lines with it \citep{Gilman1983}. \citet{Longcope1998} showed fluid motion with right handed coil deform the flux tube with a left handed.
		
		Using helioseismic measurement of sub-photospheric flows, \citet{Komm2007} have reported cyclonic vorticity (counter clockwise) in the northern hemisphere. The hemisphere preference has also been found in other studies \citep{Gao2009, Maurya2011b, Seligman2014, Komm2014a}. There are a few studies on the  combined topology of sub-photospheric flows and photospheric magnetic helicities \citep{Gao2009, Maurya2011b, Seligman2014, Komm2015b}. In some of these studies, kinetic helicity has been found to show statistically opposite HSR than that of current helicities \citep{Gao2009}. 
		
		There are not enough studies reported on the topology of the combined photospheric magnetic fields and sub-photospheric velocity fields from the peak to descending phase of the solar cycle 23. Therefore, it would be interesting to investigate the topological parameters of sub-photospheric flows, photospheric magnetic fields and their hemispheric preferences, associations. With this aim, we explore for any possible relationship between the magnetic  and kinetic helicities using photospheric vector magnetograms and Dopplergrams observations of a large sample of ARs.  Rest of the manuscript is organized as follow: In Section~\ref{sec:data-analysis}, we describe the selection of ARs and the observational data used. We discuss the methodology for computing different helicity parameters. Section~\ref{sec:result-discn} is devoted to the results and discussion, and finally, we present the summary and conclusions in Section~\ref{sec:sum-conc}.
		
		\section{The Active Regions, Data and Analysis }
		\label{sec:data-analysis}
		
		The main observational data used in this study consists of dopplergrams and vector magnetograms of several ARs of the main to decay phase of the solar cycle 23. Doppler observations of the full disc Sun for several years are available from the Global Oscillation Network Group (GONG) while the corresponding vector magnetic field observations are not available as such during the aforementioned period. Therefore, for our studies, we have obtained vector magnetograms from the Marshall Space Flight Center (MSFC). Furthermore, GONG high resolution Doppler observations are available from 22 July 2001 onward while MSFC vector magnetograms are available only for the periods 19 September 2000 to 25 October 2004. Due to this constraint, we have chosen only the common sets of data for the period of 22 July 2001 to 25 October 2004. The other limitations are the availability of daily vector magnetic field data. There are data gaps of several days. ARs far away from the disc center suffer from the foreshortening effects \citep[][ and references therein]{Maurya2014} and produce spurious results in helioseismic computations. Hence, we selected only the ARs within the radius of $\pm40^{{\rm o}}$ in heliographic longitude and latitude. In total, we short-listed 189 ARs for the common period of MSFC and GONG observations, where some of these ARs were observed during multiple rotations. In our sample of 189 ARs, 118 ARs were found to be located in the southern hemisphere and 71 ARs in the northern hemisphere. Thus the number of ARs for the southern hemisphere is larger than that in the northern hemisphere. However, we do not expect our results on HSR to be statistically affected by the different numbers of ARs in the two hemispheres.
		
		\subsection{The Vector Magnetograms and Analysis}
		\label{sec:mag-fields}
		
		The MSFC vector magnetograph facility was assembled in 1973 to support the Skylab mission, and thereafter went through several improvements \citep[MSFC,][]{Hagyard1982,Hagyard1995,West2002}. The vector magnetic fields data were taken in the magnetically sensitivity spectral line Fe \textsc{i} 5250.2\,\AA~using the Zeiss birefringent filter with band pass of 125\,m\AA. Its spatial resolution over the $7^{\prime}\times5.2^{\prime}$ field-of-view is $0.64\arcsec pixel^{-1}$. These observations are taken in the far wing of the spectral line at -90\,m\AA (or -120\,m\AA) from the line center, where the Faraday rotation effect decreases \citep{Ronan1992}. The pixel resolution of the MSFC magnetograms is 1\arcsec.28 along both x- and y-directions.
		
		The MSFC observations consist of line-of sight magnetic field ($B_{l}$), transverse component $(B_{t})$ and the azimuth angle $(\phi)$ of $B_t$. The horizontal components $(B_{x},B_{y})$ of the vector magnetic field can be computed from $ B_t $ and $ \phi $. We resolved the usual $180^{\rm\circ}$ ambiguity in the transverse components from all the vector magnetograms using the acute angle method although there are several other methods available as reviewed by \citet{Metcalf2006}. Further, to avoid any projection effects, we transformed the images to the disc center using the method described by \citet{Venkatakrishnan1989}.
		
		The MSFC vector magnetograms are not aligned in the east-west and north-south directions which should be corrected for comparing the photospheric magnetic fields with sub-photospheric flows obtained from the GONG. Active region's NOAA number and their central locations are provided in the header of every vector magnetogram. However the central coordinates are found to be incorrect in some cases. In order to correct the coordinates and orientation of the MSFC vector magnetograms, we have used the full disc SOHO/MDI magnetograms for reference. First of all, based on the location and time provided in the header of the vectors magnetograms, we estimated corresponding location in the MDI magnetograms and then after cross-correlating the ARs between MSFC and MDI magnetograms, we corrected for the locations and orientations of MSFC observations.

		\subsubsection{Determination of Magnetic and Current Helicities}
		\label{sec:determine-mag-cur-hel}
		
		As discussed in the Section~\ref{sec:intro}, the force-free parameter $\alpha_f$  (Equation~\ref{eq:alpha}) can be used as a proxy for the magnetic helicity (Equation~\ref{eq:mag-hel-def}). But, $ \mathbf{ B} $ is observed only at a single height, so we can have its derivatives only along  $ x- $ and $ y- $ directions. Thus, we can compute only the vertical part of $ \alpha_f $. Basically, there are two approaches for calculating it: \textit{i)} $(\alpha_f)_{\rm best}^{z}$ which is the best fit single value for the whole AR in the least squares sense \citep{Pevtsov1995}, and \textit{ii)} $(\alpha_f)_{av}^{z}$ which is the average of the vertical component of $\alpha_f$ over the entire AR \citep{Bao1998, Abramenko1996, Hagino2004}. Both methods have been found to be in fairly good agreement to each other \citep{Leka1999a}. For a linear force-free field, the values of $ \alpha_f $ for a given position $(x,y)$, can be given by,
		
		\begin{equation}
		\alpha_f^{z}(x,y)=\frac{\sum J_{z}\cdot B_{z}}{\sum B_{z}^{2}}=\left[\sum_{x,y}\left(\frac{\partial B_{y}}{\partial x}-\frac{\partial B_{x}}{\partial y}\right)\cdot B_{z}\right]/\sum_{x,y}B_{z}^{2}
		\label{eq:alpha-pix}
		\end{equation}
		
		\noindent where, numerical derivatives of magnetic field components can be determined using the Lagrange's three point interpolation. While computing the pixel values of $\alpha_f^{z}(x,y)$ in an AR, we selected only those pixels where the total magnetic field strength were above the quiet regions. To get the net twist in an AR, we took average over the entire AR,
		
		\begin{equation}
		(\alpha_f^{z})_{\rm av}=\frac{1}{N}\sum\limits _{x,y}\alpha(x,y)
		\label{eq:alpha-av}
		\end{equation}
		
		\noindent where, $N$ represents the total number of points where $\alpha_f^{\rm z}(x,y)$ are estimated. Along with the magnetic helicity parameter $(\alpha_f^{\rm z})_{\rm av}$, we computed the vertical  current helicity,
		
		\begin{equation}
		(h_{\rm c}^{\rm z})_{\rm av}=\frac{1}{N}\sum_{x,y}J_{z}\cdot B_{z}=\frac{1}{N}\sum_{x,y}\left(\frac{\partial B_{y}}{\partial x}-\frac{\partial B_{x}}{\partial y}\right) B_{z}
		\label{eq:cur-hel-av}
		\end{equation}

		\noindent We have also computed the net absolute magnetic flux ($ B $) for every AR by adding the fluxes for those positions where helicity is estimated.  From the above expressions for the magnetic helicity parameter $\alpha_{\rm av}^{\rm z}$ (Equation~\ref{eq:alpha-av}) and current helicity parameter $h_{\rm c}^{\rm z}$ (Equation~\ref{eq:cur-hel-av}), it is evident that both these parameters could  have same sign for a given AR. 
		
		\subsection{Sub-photospheric Flows }
		\label{sec:sub-surf-flows}
		
		The horizontal components ($u_{x}$, $u_{y}$) of the sub-photospheric flow parameters for all ARs are computed from the Doppler observations using the ring-diagram analysis \citep{Hill1988}. The basis of this method is that the frequency of sound waves inside the solar interior are perturbed by the sub-photospheric plasma motion, $ \delta\omega_{n,\ell} = \mathbf{k}_{n,\ell}\cdot\mathbf{U}_{n,\ell}$, where, $\mathbf{ k}_{n,\ell} $ is the propagation vector and $ \mathbf{U}_{n,\ell} $ is the flow vector for a given mode ($n,\ell$). Mode parameter $ \mathbf{U}_{n,\ell}$ can be computed by fitting the three dimensional Doppler power spectra \citep{Haber2000}. It is related with the sub-photospheric flows, $ \mathbf{U}_{n,\ell} = \int K_{n,\ell}(r) \mathbf{ u}(r)dr  + \epsilon_{n,\ell} $, where the kernel $ K_{n,\ell} $ can be computed based on standard solar model \citep{Christensen-Dalsgaard1996}, $ \mathbf{u}(r) $ is the horizontal velocity at radius $ r $, and $ \epsilon_{n,\ell} $ is the error in the mode parameter $ U_{n,\ell} $. This is an inverse problem, Fredholm equation of first kind, which can be solved using inversion methods such as regularized least squares \citep{Gough1985}. For more detail about the ring-diagram analysis one can refer to the review paper by \citet{Antia2007}. 
		
		The sub-photospheric flows are largely affected by the solar differential rotation and meridional circulations. We computed the residual flows by removing these large scale trend following the methods as described in \citet{Komm2004}. 
		
		Ring-diagram yields only the horizontal components, $ u_x $ and $ u_y $, of the flow, but not the vertical component, $ u_z $, at any point in the interior.  We estimated the $u_{z}$ from the divergence of horizontal components, assuming mass conservation \citep{Komm2004}, or steady-state continuity equation,
		
		\begin{equation}
			\frac{\partial\rho(z)u_z}{\partial z} + \rho(z)(\nabla\cdot\mathbf{ u})_h = 0,
			\label{eq:continuity}
		\end{equation}	
	    
	    \noindent with horizontal divergence,
	    
	    \begin{equation}
	    	div_h \equiv (\nabla\cdot\mathbf{ u})_h  \equiv \frac{\partial u_{x}}{\partial x}+\frac{\partial u_{y}}{\partial y}.
	    	\label{eq:div-h}
	    \end{equation}
	    
	    \noindent  Note that $ z $ is the vertical height increasing with radial distance with the origin on the surface. The interior is specified by a negative value of $ z $. By supposing $u_z$ on the surface, $ z=0 $, and integrating the continuity Equation~\ref{eq:continuity} over $ z $ from $ d $ (negative) to 0, we obtain,
	    
	    \begin{equation}
	    	u_z(d) = -\frac{1}{\rho(d)}\int_0^d \rho(z) (\nabla\cdot\mathbf{ u})_h dz
	    	\label{eq:uz}
	    \end{equation}
	    
	    \noindent which has the same form as Equation (5) of \citet{Komm2004}. Now, suppose $ (\nabla\cdot\mathbf{ u})_h>0 $ in the height range from $ d $ to 0, then it follows that $ u_z(d) >0 $. This means that a horizontally diverging flow rises upward, and hence $ u_z $ and $ (\nabla\cdot\mathbf{ u})_h $ are expected to have the same sign. 

		\subsubsection{Sub-photospheric Topological Parameters}
		\label{sec:sub-surf-flow-top}
		
		Following, Equation~\ref{eq:cur-hel-av} for the current helicity, we computed the vertical kinetic helicity density as follow,
		
		\begin{equation}
		h_{\rm k}^{\rm z}  = \left<u_z\cdot\omega_z\right>
		\label{eq:kin-hel-den-z}
		\end{equation}
		
		\noindent where, the vertical vorticity is given by,
		
		\begin{equation}
		\omega_{z} \equiv curl_z = \frac{\partial u_{y}}{\partial x}-\frac{\partial u_{x}}{\partial y},
		\label{eq:vert-vort}
		\end{equation}
		
		\noindent represents the twist of the flow around the vertical (radial direction). The parameter $ h_{\rm k}^{\rm z}$ is analogous to $ h_{\rm c}$ (Equation~\ref{eq:cur-hel-den}), and we have taken it as a proxy for the kinetic helicity density.

		Note that the  vertical kinetic helicity density $h_{\rm k}^{z}$ (Equation~\ref{eq:kin-hel-den-z}) is obtained from the weighted average of $\omega_{z}$ by $ u_z $. But one should note that in strong magnetic field regions vertical flows are suppressed due to the Lorentz force. The horizontal divergence (Equation~\ref{eq:div-h}) of flows from the sunspot regions to outwards are common. Therefore, weighted average of  $\omega_{z}$ with $ div_h $ would be a better estimate for the vertical helicity. Such a parameter can be given by,
		
		\begin{equation}
		C_{\rm k}^{\rm z} = \left<div_h\cdot curl_z\right>
		\label{eq:div-curl-z}
		\end{equation}
		
		The parameter, $ C_{\rm k}^z $, is expected to have same sign as that of $ h_{\rm k}^z$ because $ div_h $ and $ u_z $ are expected to have same sign as explained earlier. This expression for the vertical twist has been used earlier for studying the kinetic helicity of supergrannular flows \citep{Egorov2004} and sub-photospheric flows \citep{Komm2007}. The validity of above helicity proxy has been confirmed for the depth range obtained from ring-diagram as numerically established by \citet{Miesch2008}.
		
		In order to compute the sub-photospheric flow topological parameters for an AR, we first computed them over patches of $ 60\deg\times60\deg $ around the disc center. Then the parameters associated with the ARs were interpolated based on their coordinates on the solar disc.  
		
		Earlier studies \citep{Komm2005a,Maurya2009b,Maurya2010d} have shown bipolar structure of sub-photospheric flows in several ARs. Therefore, two different types of twists are possible in the underlying flows of ARs. To deal with it, we analysed $ h_k^z $ and $ C_k^z $ at the two depths of 2.4\,Mm and 12.6\,Mm. For distribution, the associated parameters are labeled by the subscripts `1' and `2', \eg, $ h_{k1}^z $ and $ h_{k2}^z $, respectively. 
		
		The sub-photospheric flows of ARs are obtained from the 1664 minute time-series duration with a cadence of one-minute Doppler observations and thus correspond to average subsurface flows over a day. Therefore, for comparing magnetic field topology with sub-photospheric flows of an AR, the magnetic field topological parameters are also averaged accordingly.
		
		Note that, next section onward, we have used the vertically averaged quantities only, and therefore, we have dropped the scripts ``z'' and ``av'' from all the topological parameters, \eg, we have used $ h_{\rm c} $ instead of $ (h_{\rm c})_{\rm av}^{\rm z} $.

		\section{Results and Discussion}
		\label{sec:result-discn}
		
		Results of our analysis of the 189 ARs are shown in Figures \ref{fig:mag-cur-hel-lat} -- \ref{fig:correl-kin-mag-hel-depth}. The hemispheric distribution of topological parameters are summarized in the Tables~\ref{tab:hemisp-distb} and~\ref{tab:top-para-bav}.
		
		\subsection{Photospheric Magnetic Fields}
		\label{sec:phot-mag-fields}
		
		The photospheric magnetic field topology is studied in terms of  magnetic helicity parameter $\alpha_f$~(Section~\ref{sec:mag-hel}) and current helicity   $h_c$~(Section~\ref{sec:cur-hel}). In the following, we discuss the results obtained.
		
		\subsubsection{Magnetic Helicity}
		\label{sec:mag-hel}
		
\begin{figure}
			\centering 
			\noindent \includegraphics[width=0.49\textwidth, clip, viewport=26 6 350 276]{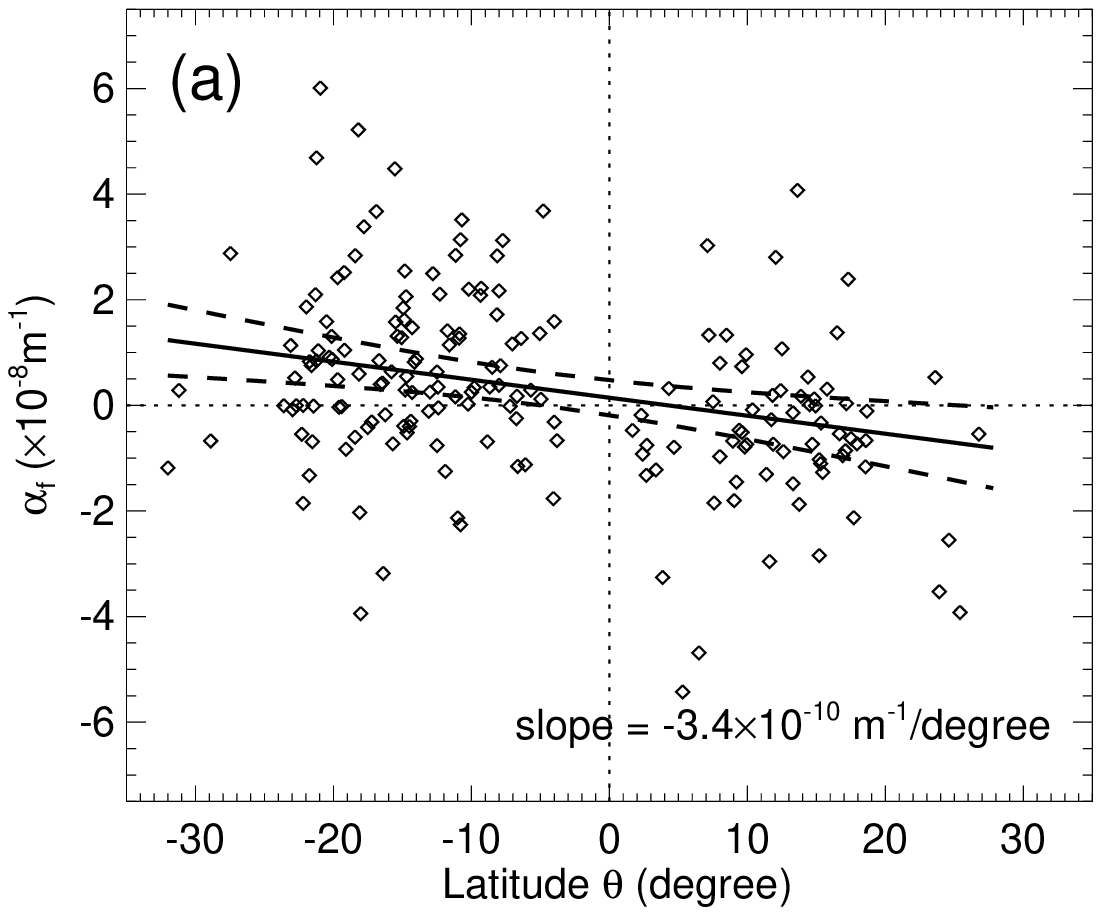} 
			\includegraphics[width=0.49\textwidth, clip, viewport=23 5 348 274]{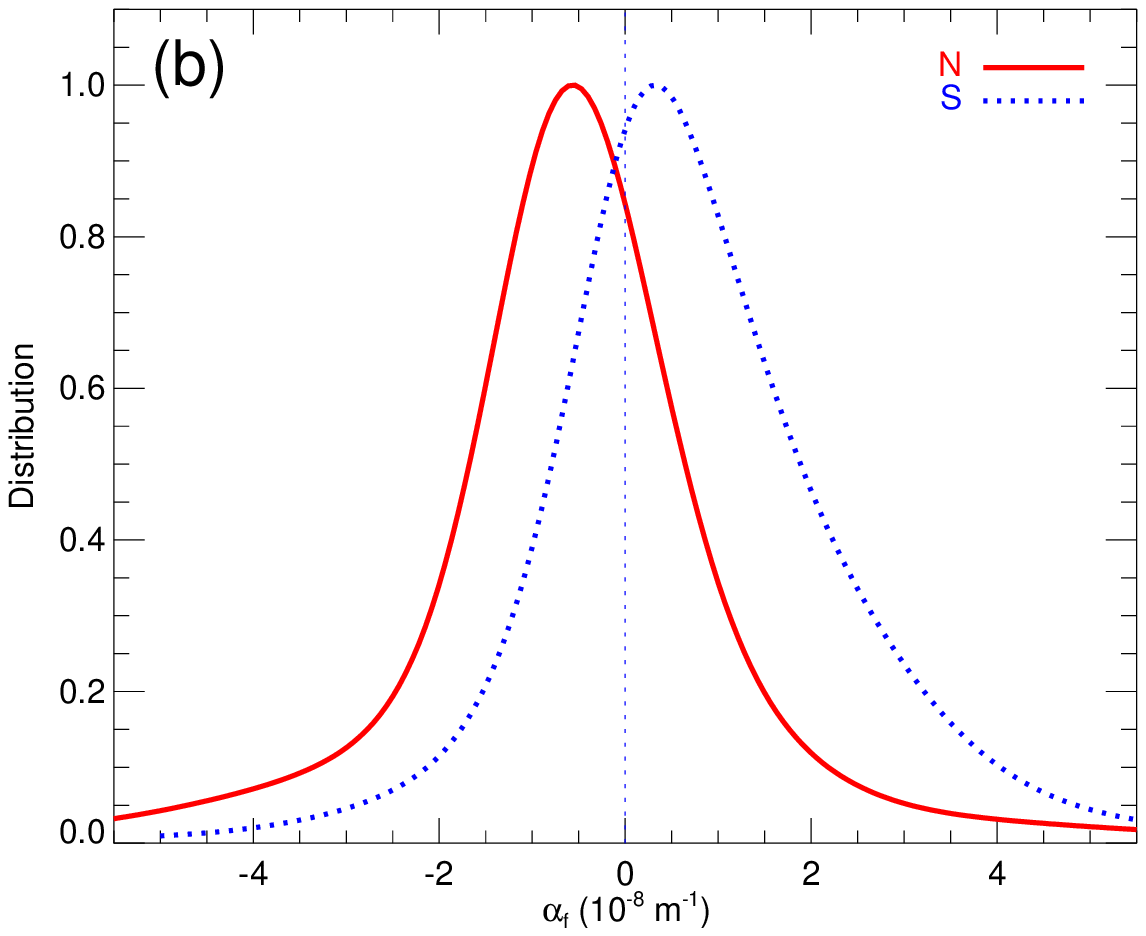}\\
			\includegraphics[width=0.49\textwidth,clip,viewport=16 6 350 276]{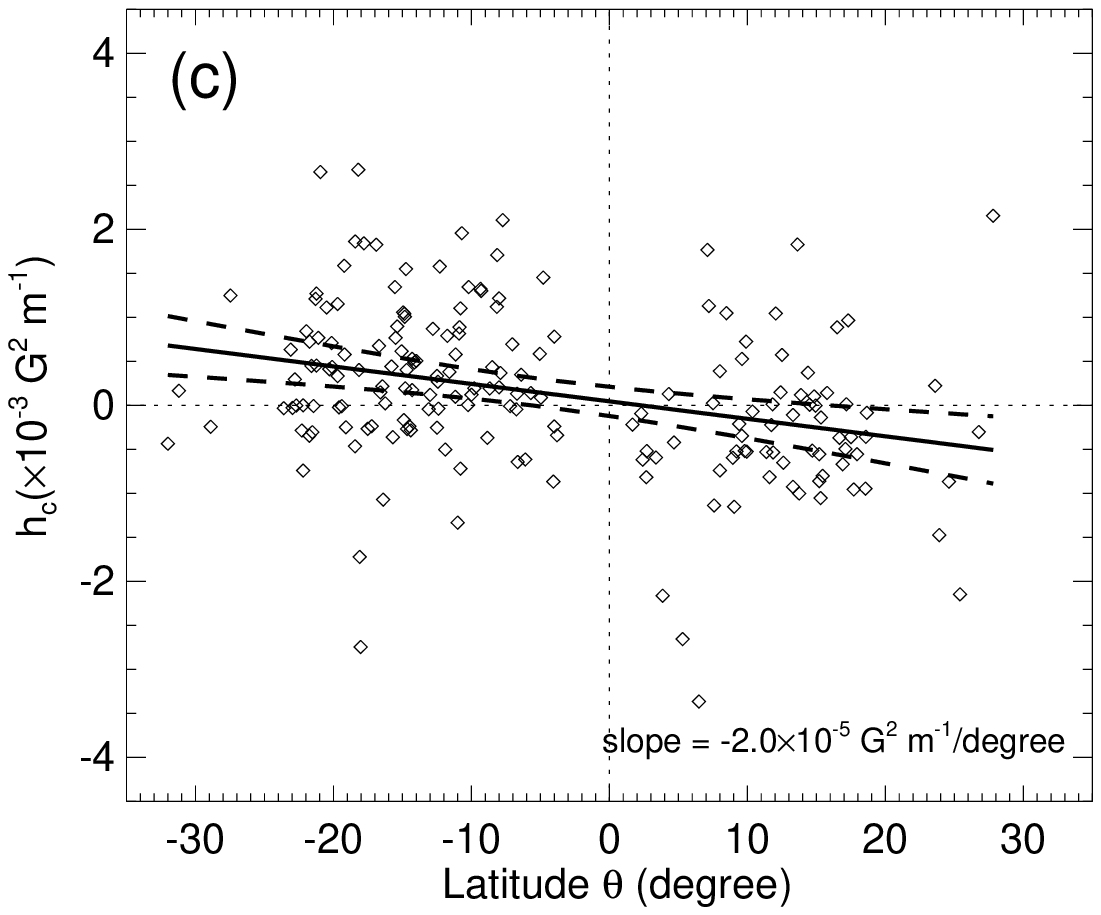} 
			\includegraphics[width=0.49\textwidth,clip,viewport=17 5 348 274]{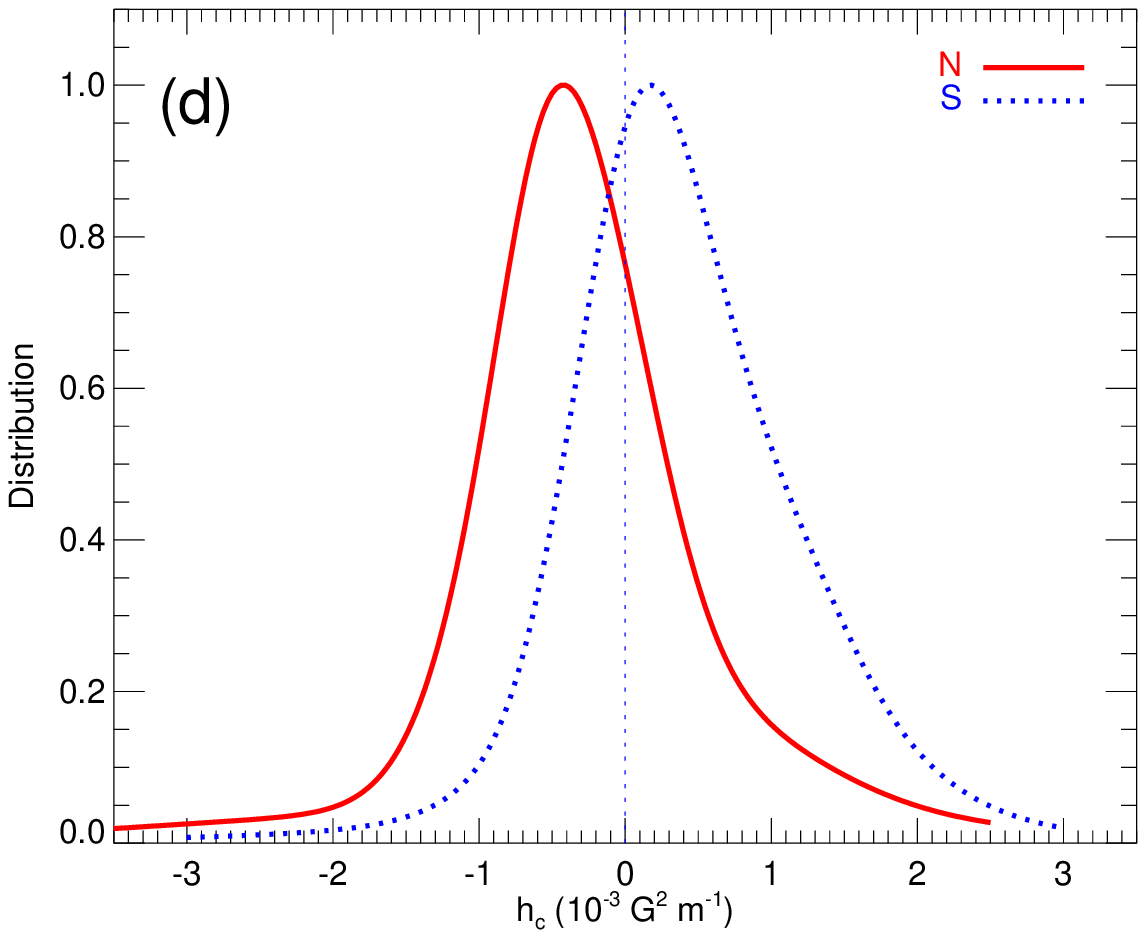}
			\caption{(a) Latitudinal distribution of the force-free parameter $\alpha_f$, where solid lines represent the linear least square fitting between the parameter $\alpha_f$~and latitude, while the dashed curves correspond to 95$\%$ confidence intervals. (b) Probability distribution function of the parameter $\alpha_f$~for the northern hemisphere (solid \& red) and southern hemisphere (dotted \& blue). Bottom row is similar to the top row but for the current helicity $ h_c $.}
			\label{fig:mag-cur-hel-lat} 
		\end{figure}

\begin{figure}
			\centering 	
			\includegraphics[width=1.0\textwidth,clip,viewport=38 3 508 242]{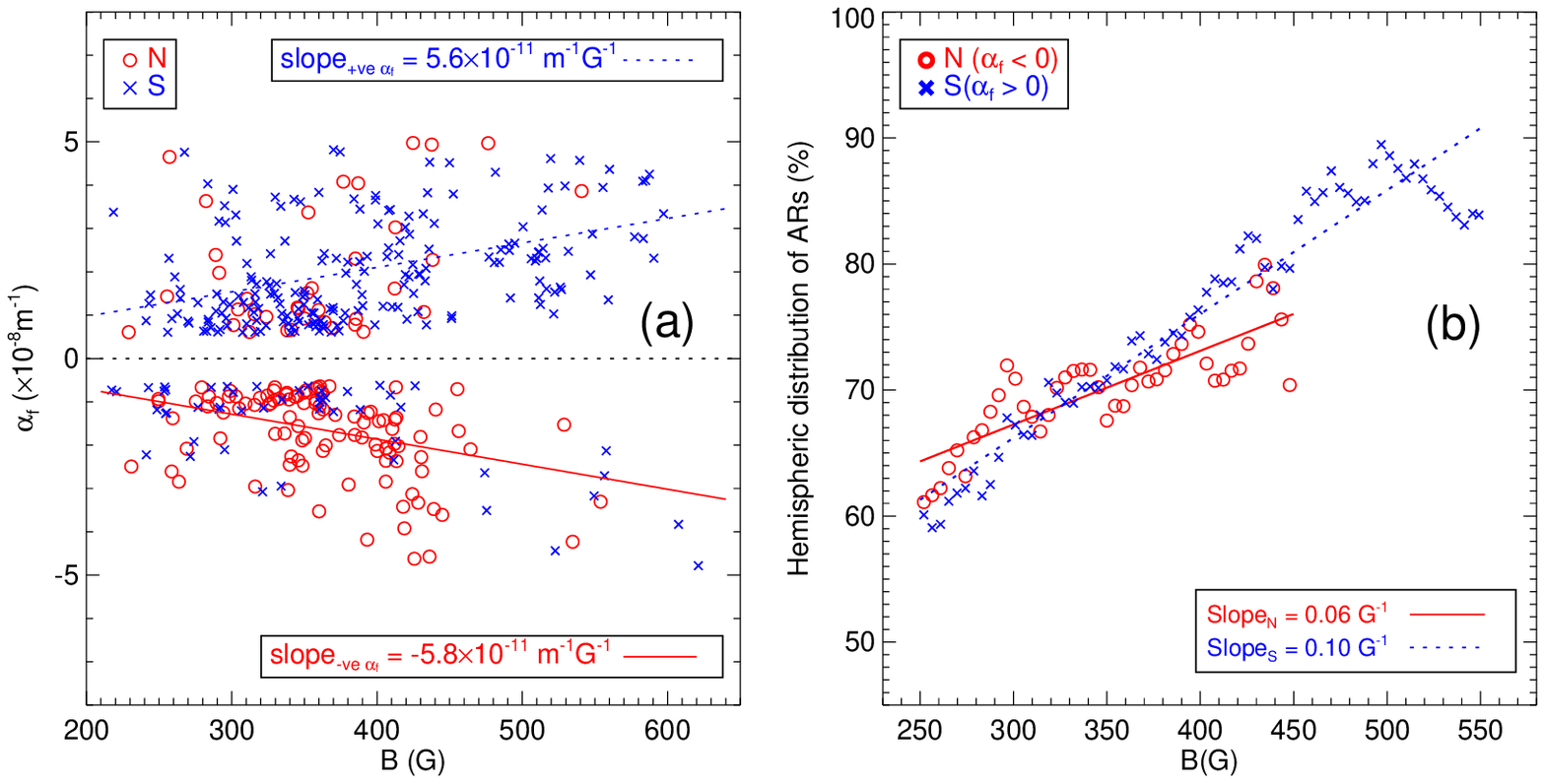}\\	
			\includegraphics[width=1.0\textwidth,clip,viewport=38 3 510 242]{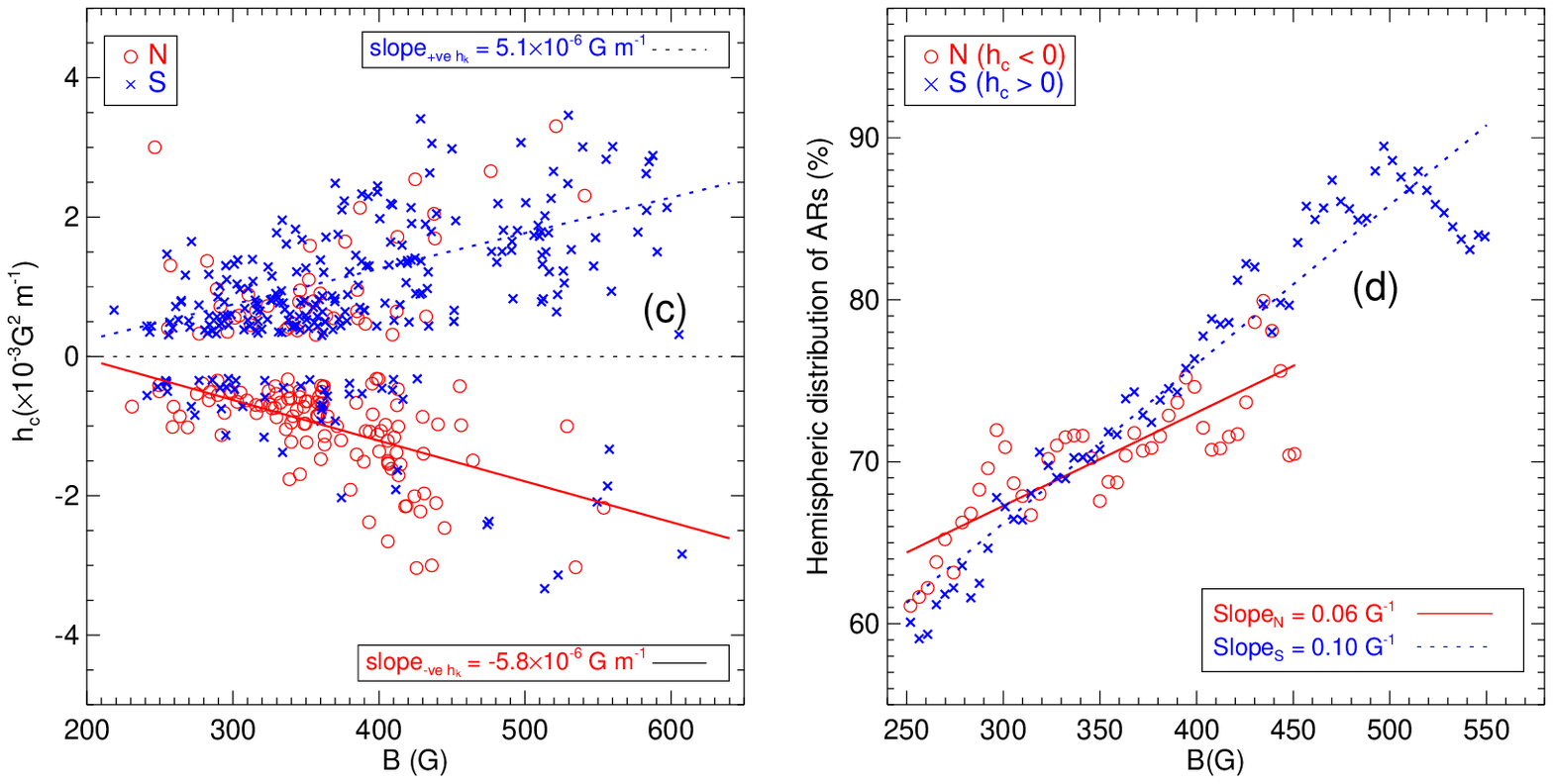}
			\caption{(a) The helicity parameter $\alpha_f$~as a function of absolute magnetic flux of ARs.  Solid (dashed) line below (above) $\alpha_f=0$ represents the linear least square fitting through northern (southern) data points. The slopes of the fitted lines are mentioned within the rectangular boxes. (b) Hemispheric distribution of ARs based on $ \alpha_f $ values as a function of magnetic field. The dashed (solid) lines represent the linear regression for the northern (southern) ARs. Corresponding equations are given in the rectangular boxes.  Bottom row is similar to the top row but for the current helicity $ h_c $.}
			\label{fig:mag-hel-absB}
		\end{figure}

\begin{figure}
			\centering
			\includegraphics[width=0.8\textwidth,clip,viewport=15 5 420 168]{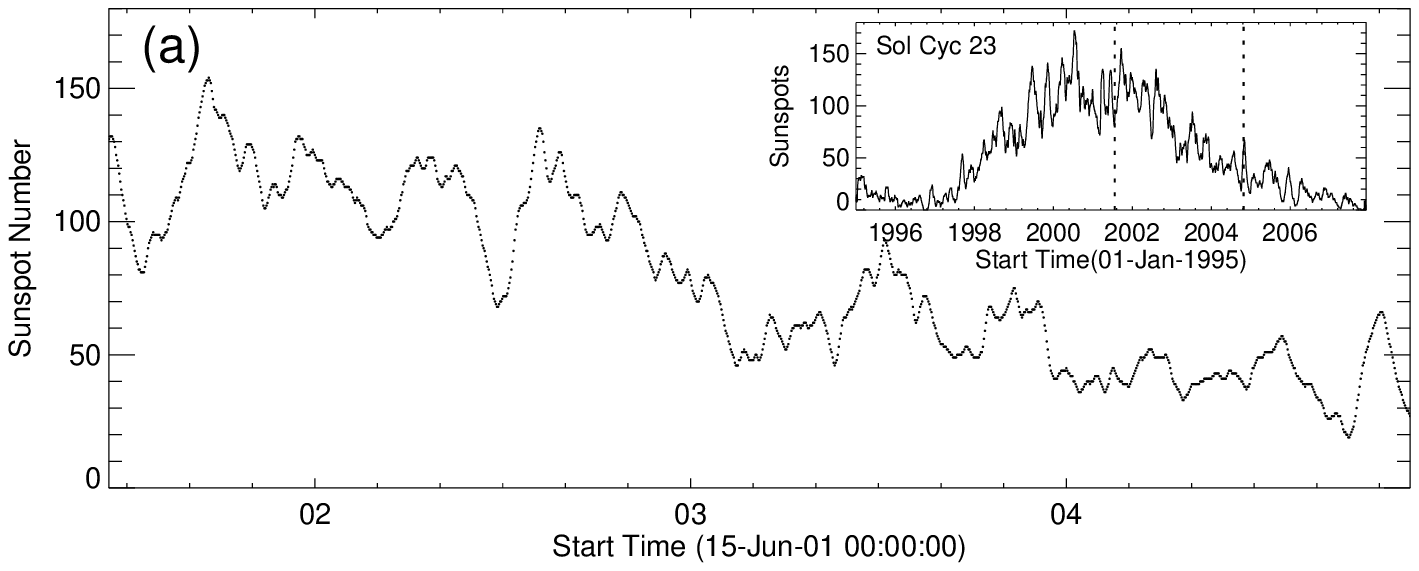}\\
			\includegraphics[width=0.8\textwidth,clip,viewport=15 5 420 240]{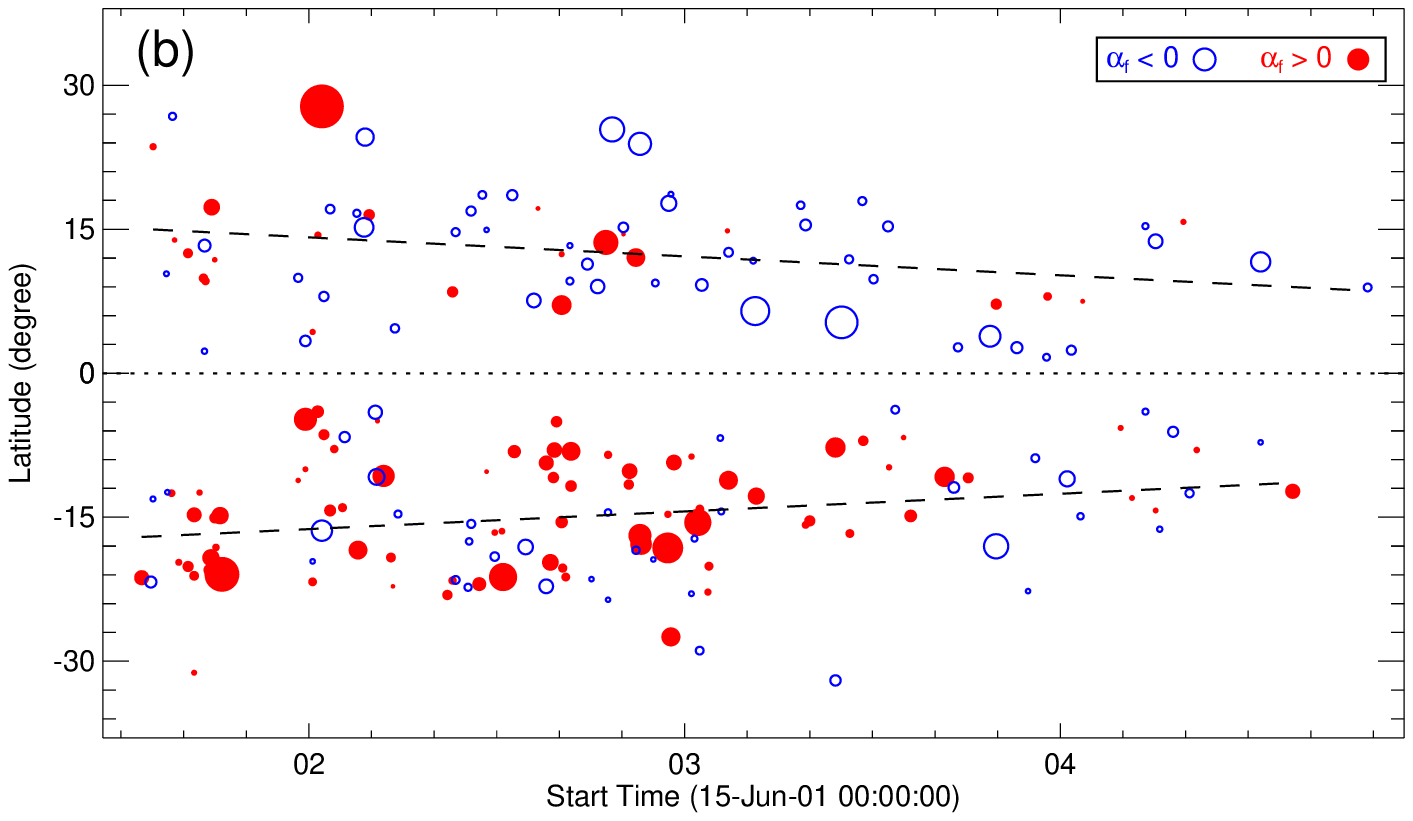}\\	
			\includegraphics[width=0.8\textwidth,clip,viewport=15 5 420 240]{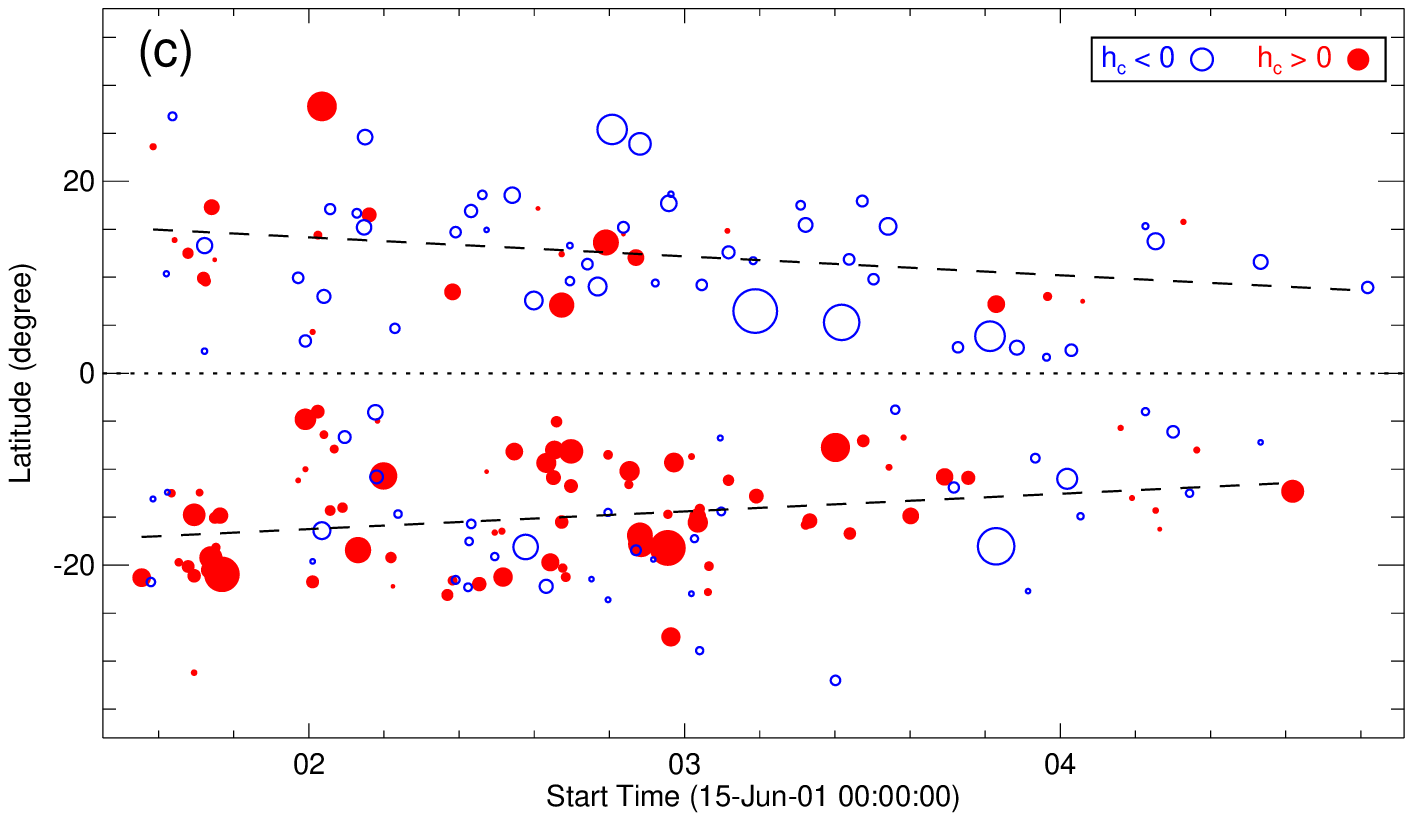}\\ 
			\includegraphics[width=0.8\textwidth,clip,viewport=15 6 420 168]{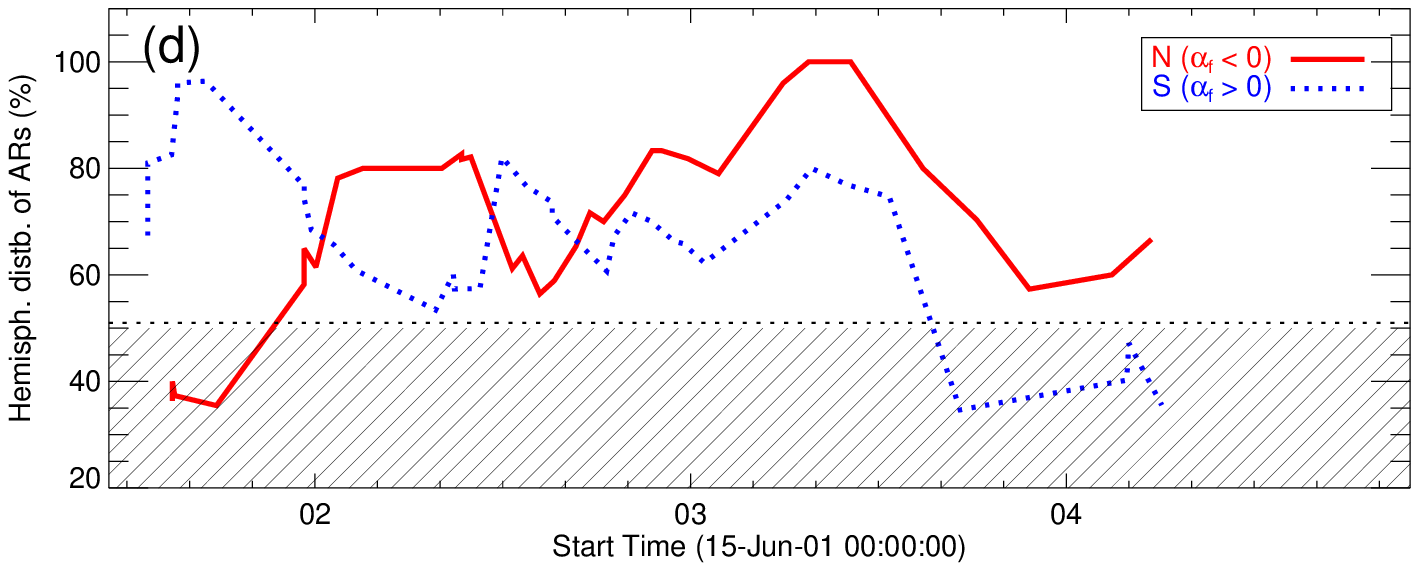}\\
			\caption{(a) Sunspot number as a function of time for reference. The complete sunspot cycle 23 is shown in the inset, where vertically dashed lines mark the time period of ARs which are analyzed in this paper. (b) Latitudinal distribution of magnetic helicity parameter $\alpha_f$~as a function of time, where sizes of opened(filled) circles represent magnitudes of negative (positive) values of the parameter $\alpha_f$, and dashed lines  represent the linear regression lines through ARs latitude. (c) Similar to the panel (b) but for the current helicity parameter $ h_c $. (d) Hemispheric distribution of ARs as a function of time for northern (solid) and southern (dotted) hemispheres.}
			\label{fig:mag-cur-hel-time}
		\end{figure}

		The results of our analysis for the magnetic helicity parameter $ \alpha_f $ are shown in Figures~\ref{fig:mag-cur-hel-lat} --~\ref{fig:mag-cur-hel-time}. Figure~\ref{fig:mag-cur-hel-lat}(a) shows the latitudinal distribution of magnetic helicity parameter $\alpha_f$~for the 189 ARs. We found $68\%(67\%)$ ARs in the northern (southern) hemisphere with negative (positive) $\alpha_f$. This result further confirms the earlier reports on HSR for the magnetic helicity. We notice that $\alpha_f$~shows large scatter in both the hemispheres. Therefore, the following questions arise: Is the hemispheric sign statistically significant? How do $\alpha_f$~values vary with latitude?
		
		To address the above questions, we fitted a linear regression model through the parameters $\alpha_f$. In Figure~\ref{fig:mag-cur-hel-lat}(a), the solid line shows the linear regression line while dashed curves correspond to the $95\%$ confidence ($\sim2\sigma$) interval. The confidence interval $95\%$ contains the best-fit regression line. The fitted line has a slope, $-3.4\times10^{-10}m^{-1} degree^{-1} $, such that the magnitude of $\alpha_f$~increases with latitude. However, linear regression is not fully satisfactory because we do not have any known reason for the straight line to be the appropriate fit. But at the same time, we can not justify higher order polynomials. The average values of $ \alpha_f $ in northern (southern) hemisphere is found to be $ -4.7 (+7.6)\times10^{-9}m^{-1}$.
		
		To examine the statistical significance of hemispheric trend, we calculated probability distribution function (PDF) for $\alpha_f$~which is shown in Figure~\ref{fig:mag-cur-hel-lat}(b). The peak of density function for northern (southern) hemisphere lies at \hbox{$-5.4(+3.4)\times10^{-8}m^{-1}$}. These results confirm the hemispheric trend of the estimated $\alpha_f$~values for our set of ARs. The FWHM of PDF of $\alpha_f$~for northern (southern) hemispheric points is \mbox{$2.3(2.6)\times10^{-8}m^{-1}$}. That is, $\alpha_f$~of the southern hemisphere is spread slightly more than that of the northern hemisphere. 
		
		Next we address whether ARs of different magnetic field strengths have similar twist in their field lines. For this, we plotted the magnetic field parameter $\alpha_f$~as a function of absolute magnetic flux ($ B $) in Figure~\ref{fig:mag-hel-absB}(a). In this plot, we have only shown the data points with $|\alpha_f|\geq0.5\times10^{-8}{\rm m^{-1}}$ for clarity. A linear regression is performed through positive and negative values of $ \alpha_f $ of ARs. The values of slope of the regression is found to be $ +5.6(-5.8)\times10^{-11} m^{-1}G^{-1}$ for positive (negative) $ \alpha_f $. This result shows that the ARs with stronger magnetic fields have larger twist irrespective of ARs's hemispheric locations.
		
		Figure~\ref{fig:mag-hel-absB}(b) shows magnetic field related HSR in the helicity parameter $ \alpha_f $. For this, we ran a boxcar magnetic flux of size 50\,G, and counted ARs following the HSR as a function of $ B $. Linear regressions through the data are shown with dashed lines with a slope $0.06(0.10)$. This implies that the HSR becomes stronger with magnetic fields strength of ARs.

		Next, we explore the time dependent hemispheric trend of the magnetic helicity parameter $ \alpha_f $ (Figure~\ref{fig:mag-cur-hel-time}). Figure~\ref{fig:mag-cur-hel-time}(a) shows the monthly averaged sunspot number as a function of time; in the period where ARs are taken for this paper. A complete solar cycle 23 is also shown in the inset of the top panel where vertically dotted lines mark the time window over which ARs are taken for this analysis. This period corresponds to the peak to the descending phase of the solar cycle 23. 
		
		Magnetic helicity parameter $\alpha_f$~for the northern and southern hemisphere is plotted in  Figure~\ref{fig:mag-cur-hel-time}(b). The symbols with opened(filled) circles represent negative(positive) values of the magnetic helicity parameter $\alpha_f$. This panel not only shows the predominantly hemispheric distribution of the parameter $\alpha_f$~but also part of the butterfly-diagrams for sunspots which indicates that the sunspots appear at larger latitude in the beginning of solar cycle and as time passes they appear at lower latitude as shown by the linear regression lines (dashed) through ARs latitudes.
		
		\begin{table}
			\begin{tabular}{|c|c|ccr|rr|rc|}
\hline
Parameter (P)                      &  N/S &    HSR &       HSR&     Total&                    Slope &        Average &            PDF &            PDF\\
                                   &      &   (\%) &     (no.)&     (no.)&$(\times10^{-2}deg^{-1})$ &                &           Max. &           FWHM\\
\hline
$\alpha_f$                         &    N &     68 &     48 &     71 &                     -3.4 &           -0.5 &           -0.5 &            2.3\\
$(m^{-1})$                         &    S &     67 &     79 &    118 &                          &            0.8 &            0.3 &            2.6\\
\hline
$h_c$                              &    N &     68 &     48 &     71 &                     -2.0 &           -0.3 &           -0.4 &            1.3\\
$(G^2m^{-1})$                      &    S &     68 &     80 &    118 &                          &            0.4 &            0.2 &            1.5\\
\hline
$h_{k1}$                           &    N &     65 &     46 &     71 &                     -1.8 &           -0.3 &           -0.2 &            2.0\\
$(ms^{-2})$                        &    S &     56 &     66 &    118 &                          &            0.2 &            0.1 &            1.6\\
\hline
$h_{k2}$                           &    N &     69 &     49 &     71 &                    -12.4 &           -1.9 &           -0.7 &            8.1\\
$(ms^{-2})$                        &    S &     67 &     79 &    118 &                          &            1.7 &            1.1 &            9.8\\
\hline
$C_{k1}$                           &    N &     65 &     46 &     71 &                     -2.4 &           -0.4 &           -0.3 &            2.2\\
$(s^{-2})$                         &    S &     55 &     65 &    118 &                          &            0.2 &            0.0 &            1.9\\
\hline
$C_{k2}$                           &    N &     65 &     46 &     71 &                     -2.5 &           -0.4 &           -0.2 &            1.8\\
$(s^{-2})$                         &    S &     70 &     83 &    118 &                          &            0.3 &            0.4 &            2.2\\
\hline
\end{tabular}

			\caption{Himispheric distribution of topolgical parameters. For the actual values of the parameters listed in the columns 6-9, the multiplication factors, $ 10^{-8}, 10^{-3}, 10^{-9} $ and $ 10^{-15} $ should be applied for $ \alpha_f $, $ h_c $, $ h_k $ and $ C_k $, respectively.}
			\label{tab:hemisp-distb}
		\end{table}
	
		\begin{SCtable}[][h]
			\begin{tabular}{|l|cr|rr|}
\hline
Parameter                          &                    +ve P&                    -ve P&                    HSR-N&                    HSR-S\\
    (P)  & \multicolumn{2}{c|}{($G^{-1}$)} & \multicolumn{2}{c|}{($\times10G^{-1}$)}\\
\hline
$\alpha_f(m^{-1})$                 &                      5.6&                     -5.8&                      0.6&                      1.0\\
$h_c(G^2m^{-1})$                   &                      5.1&                     -5.8&                      0.6&                      1.0\\
\hline
$h_{k1}(ms^{-2})$                  &                      0.1&                     -0.1&                      1.9&                      1.1\\
$h_{k2}(ms^{-2})$                  &                      1.4&                     -1.3&                     -1.1&                     -1.2\\
\hline
$C_{k1}(s^{-2})$                   &                      0.8&                     -3.0&                      0.9&                      1.2\\
$C_{k2}(s^{-2})$                   &                      1.6&                     -2.6&                     -0.7&                     -0.9\\
\hline
\end{tabular}

			\caption{Slopes of the distributions of the topological parameters (columns $ 2^{nd} $ and $ 3^{rd} $) and their hemispheric distributions (columns $ 4^{rth} $ and $ 5^{th} $) as a function of $ B $. For the actual values of the parameters listed in the columns 1-2, the multiplication factors, $ 10^{-8}, 10^{-3}, 10^{-9} $ and $ 10^{-15} $ should be applied for $ \alpha_f $, $ h_c $, $ h_k $ and $ C_k $, respectively.}
			\label{tab:top-para-bav}
		\end{SCtable}	
	
		These distributions show a clear indication of the hemispheric distribution of the magnetic helicity in both the northern and southern hemispheres. Around the end of year 2002, there are ARs with slightly large magnetic helicity. During the end phase of the solar cycle, the magnetic helicity reduces, as is obvious from the smaller circles. Overall, there seems to be equator-ward propagation of the magnetic helicity parameter $ \alpha_f $ similar to the sunspots of the solar cycle. Such results have also been reported in another study by \citet{Zhang2010} in their data obtained from other sources. 
		
		In order to get the time dependent hemisphere distribution of the parameter $ \alpha_f $, we ran a boxcar of width 90 days and counted the numbers of ARs following the HSR in both the hemispheres. Figure~\ref{fig:mag-cur-hel-time}(d) shows the boxcar averaged hemispheric distribution of the parameter $ \alpha_f $. This plot shows that the ARs on average follow HSR for the magnetic helicity parameter $ \alpha_f $ except at peak and end phase of the cycle for northern and southern ARs, respectively. The opposite hemispheric trend during the decay phase of this solar cycle has been found in earlier studies \citep[\eg,][]{Bao2000,Tiwari2009a,Pipin2019}.

		\subsubsection{Current Helicity}
		\label{sec:cur-hel}
		
		Next, we analysed the current helicities $ h_c $ of all the 189 ARs similar to the magnetic helicity parameter $ \alpha_f $ (see Section~\ref{sec:mag-hel}). The results of our analysis are shown in Figures~~\ref{fig:mag-cur-hel-lat} --~\ref{fig:mag-cur-hel-time}. Figure~\ref{fig:mag-cur-hel-lat}(c) shows the latitudinal distribution of the vertical current helicity $h_c$. There are 68\% ARs in both northern and southern hemispheres with negative and positive current helicities $ h_c $, respectively. The small deviation in $ h_c $ from $ \alpha_f $ for the southern ARs is due to averaging of $ h_c $ values for ARs that are observed multiple times.  The values of $h_c$~show a similar hemispheric trend as the force-free parameter $\alpha_f$~(Figure~\ref{fig:mag-cur-hel-lat}a). The slope of linear least square fit through $ h_c $ of ARs is found to be $ -2.0\times10^{-10} G m^{-1} degree^{-1} $.  This hemispheric trend of ARs based on $ h_c $ confirms the earlier reports \citep[][and others]{Abramenko1996}.
		
		The probability distribution function for $h_c$ is shown in Figure~\ref{fig:mag-cur-hel-lat}(d). This shows hemispheric trend for $h_c$ as the PDFs for northern and southern hemispheric ARs peak at $-4.1\times10^{-4}G^2m^{-1}$ and $+1.8\times10^{-4}G^2m^{-1}$, respectively. The FWHM of PDF for northern hemisphere is $1.3\times10^{-3} {\rm G}^2\,{\rm m}^{-1}$ and for the southern hemisphere $1.5\times10^{-3} {\rm G}^2\,{\rm m}^{-1}$. The PDF profiles of the parameter $ h_c $ and the parameter $\alpha_f$ are look-alike  representing similar distributions of two topological parameters in ARs.

		Figure~\ref{fig:mag-hel-absB}(c) shows the variation of current helicity $h_c$~as a function of absolute magnetic flux ($B$). It is found that $h_c$~increases on increasing the magnetic flux of ARs similar to the magnetic helicity parameter $\alpha_f$~(Figure~\ref{fig:mag-hel-absB}a). The slope of the linear least square fitting is found to be $ 5.1(-5.8)\times10^{-6} G m^{-1}$ for the positive (negative) $ h_c $ values of ARs. This suggests that the ARs with stronger magnetic fields may have more current helicity than typical ARs. Figure~\ref{fig:mag-hel-absB}(d) shows the hemispheric distribution of the current helicity $ h_c $ as a function of $ B $, similar to Figure~\ref{fig:mag-hel-absB}(b) for the helicity parameter $ \alpha_f $. It shows that the HSR becomes stronger with the magnetic field strength $ B $ as the slopes for northern(southern) ARs is found to be $0.06(0.10)$.

		We have also analyzed the solar cyclic variation in the current helicity $h_c$. Figure~\ref{fig:mag-cur-hel-time}(c) shows the hemispheric distribution of the current helicity parameter $h_c$~as a function of time.  This plot further confirms the equator-ward propagation of helicity $ h_c $ patterns.  Since $ h_c $ has same sign as $ \alpha_f $, it shows similar time dependent distribution as shown in  Figure~\ref{fig:mag-cur-hel-time}(d) for the parameter $ \alpha_f $.

		\begin{figure}[H]
			\centering	
			\includegraphics[width=0.24\textwidth,clip,viewport=7 12 242 242]{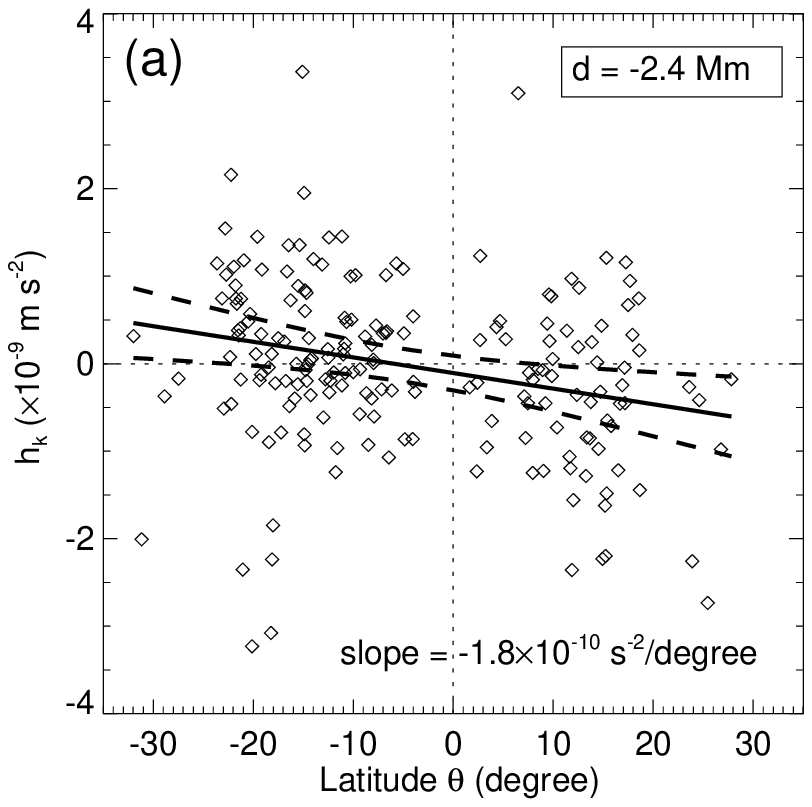}
			\includegraphics[width=0.24\textwidth,clip,viewport=7 8 242 242]{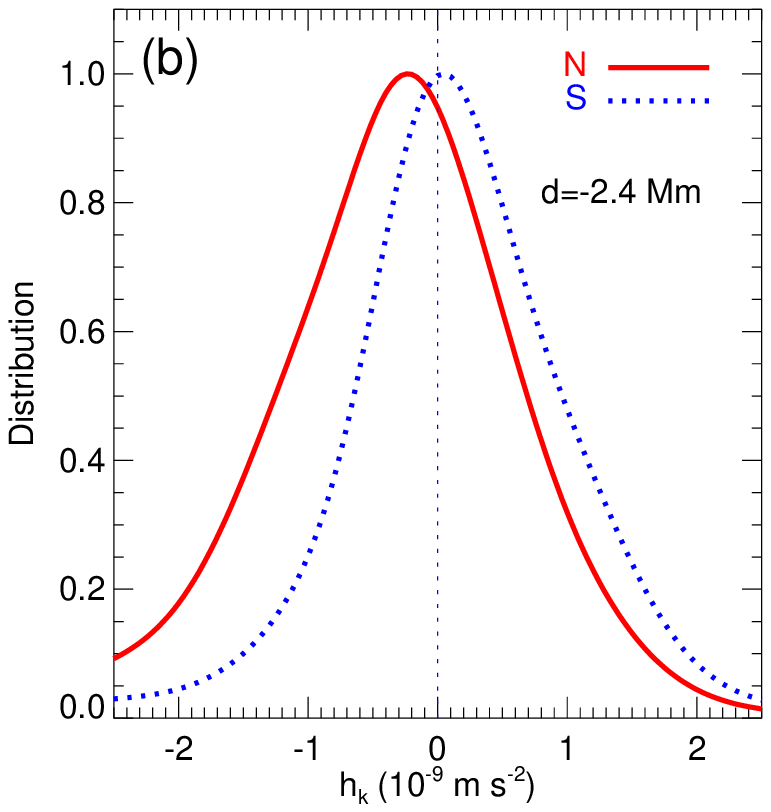}
			\includegraphics[width=0.24\textwidth, clip,viewport=7 12 242 242]{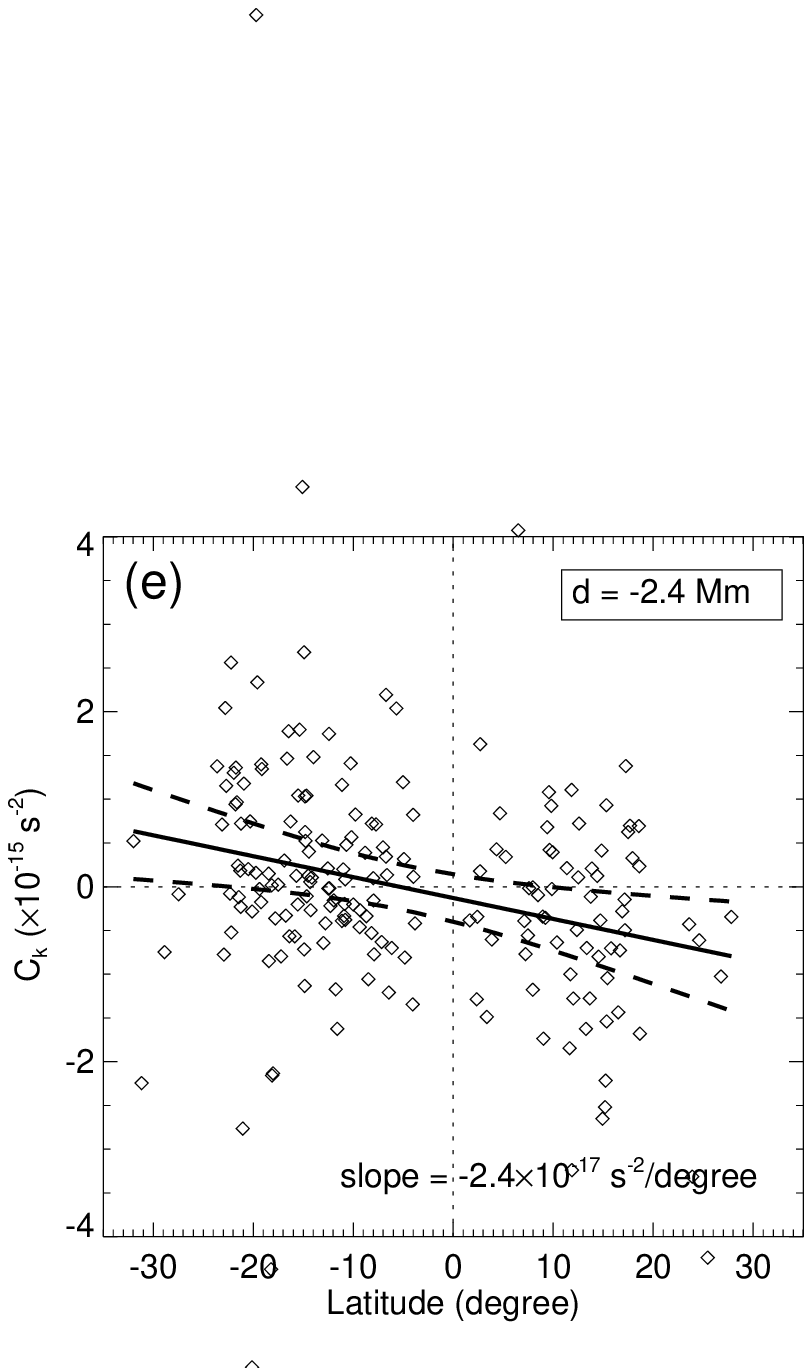}
			\includegraphics[width=0.24\textwidth, clip,viewport=7 8 242 242]{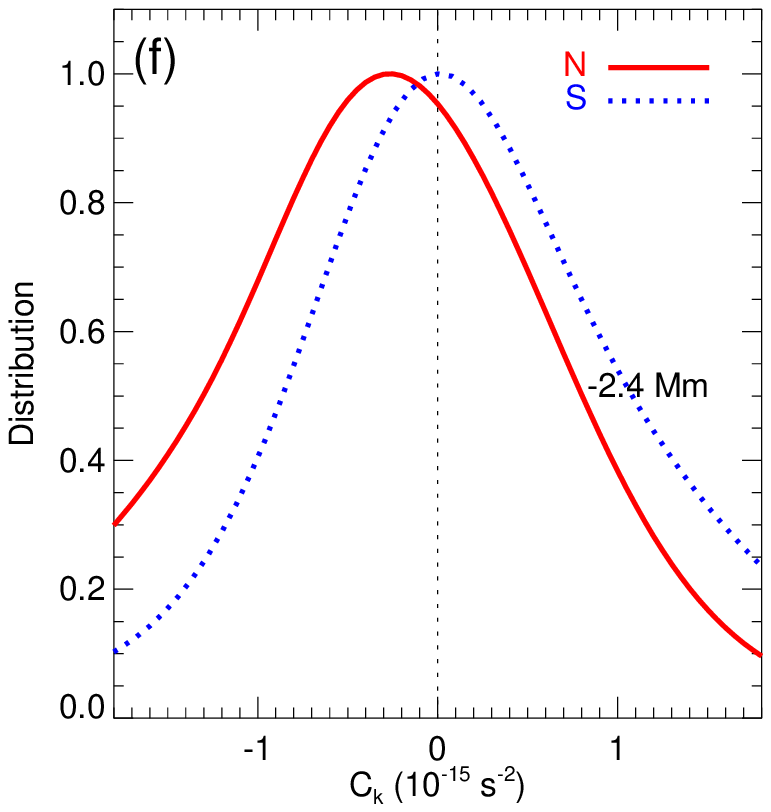}\\
			\includegraphics[width=0.24\textwidth,clip,viewport=7 12 242 242]{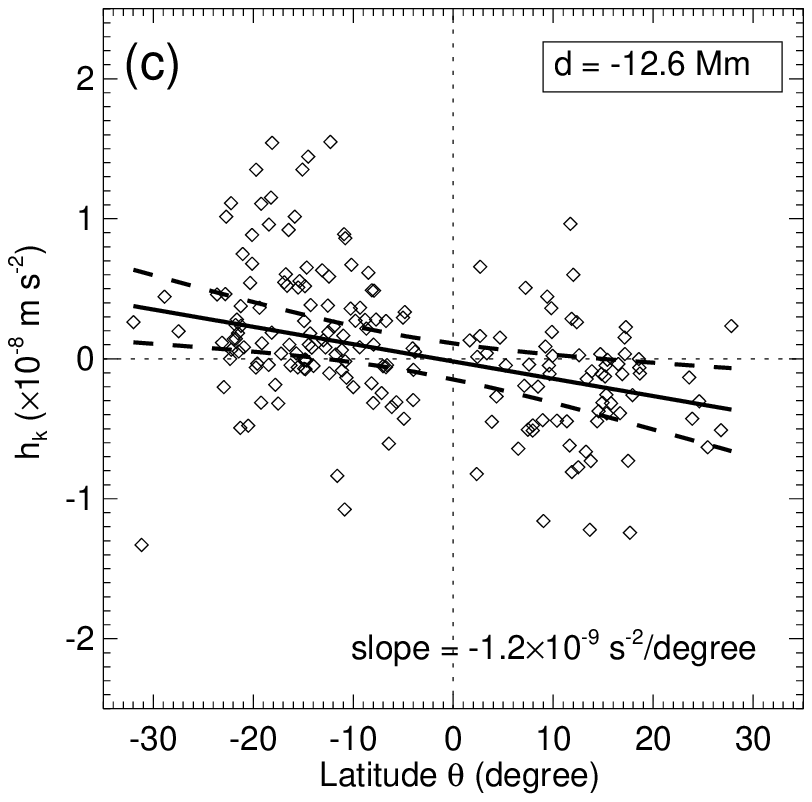} 
			\includegraphics[width=0.24\textwidth,clip,viewport=7 8 242 242]{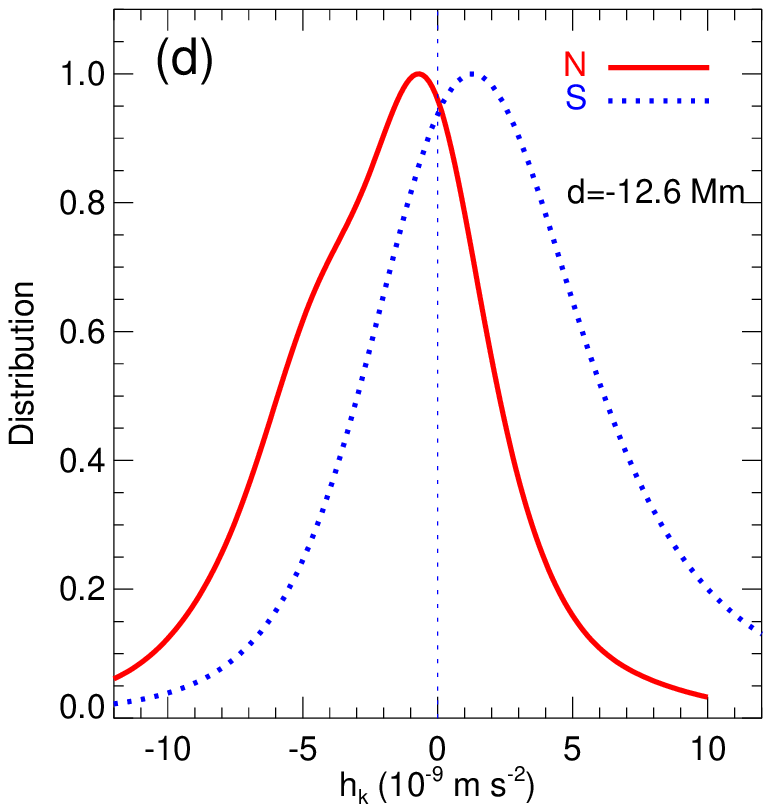} 
			\includegraphics[width=0.24\textwidth, clip,viewport=7 12 242 242]{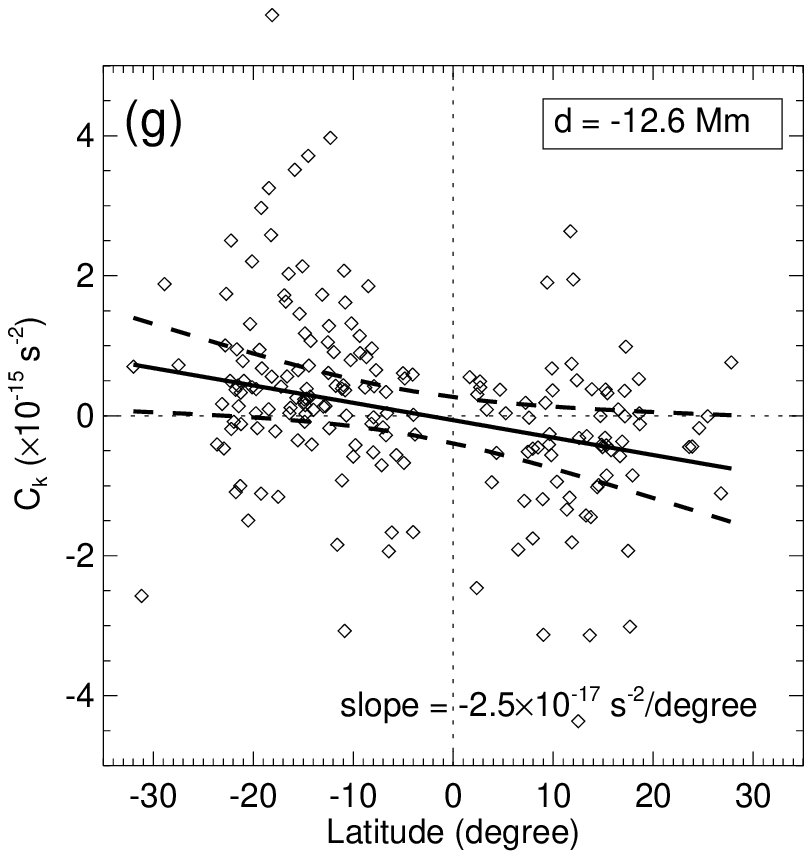}
			\includegraphics[width=0.24\textwidth, clip,viewport=7 8 242 242]{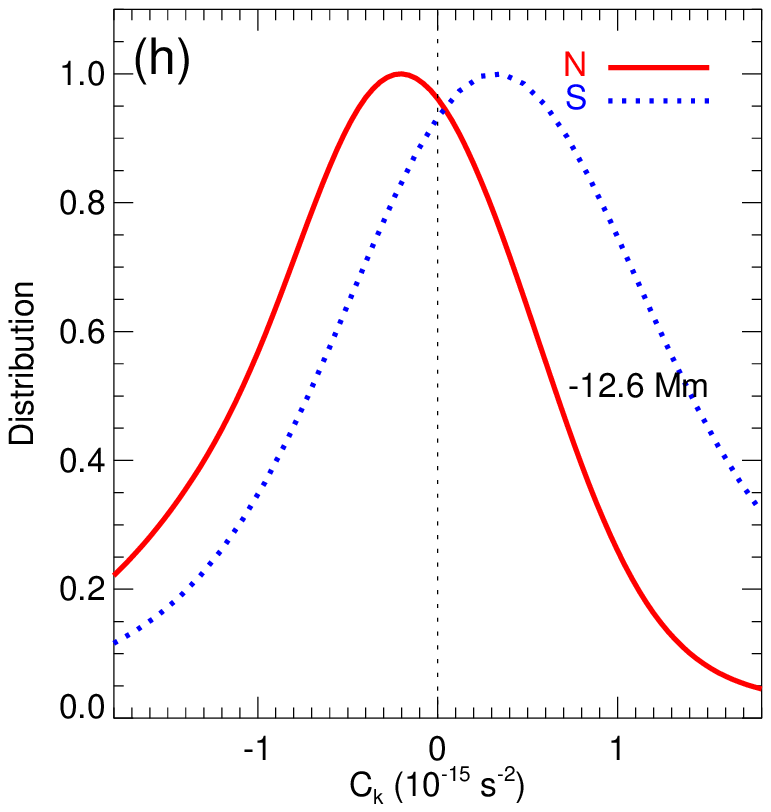}	
			\caption{Similar to Figure~\ref{fig:mag-cur-hel-lat} but for the kinetic helicity density $h_k$ (first two columns) and divergence-curl $ C_k $ (last two columns) at two different depths 2.4\,Mm (top row) and  12.6\,Mm (bottom row).}
			\label{fig:kin-hel-lat-pdf}
		\end{figure}

		\subsection{Sub-photospheric Flow Fields}
		\label{sbsec:subsurf-flow}
		
		Helicities in the sub-photospheric velocity fields in our sample of 189 ARs are studied by computing the kinetic helicity density $h_k$ (Equation~\ref{eq:kin-hel-den-z}) and the divergence-curl $C_k$ (Equation~\ref{eq:div-curl-z}). In Figures~\ref{fig:kin-hel-lat-pdf} --~\ref{fig:kin-hel-time-d2}, we explain our results based on these parameters.
			
		\subsubsection{Kinetic Helicity Density}
		\label{sec:kin-hel}
		
		Figure~\ref{fig:kin-hel-lat-pdf}(a-d) shows the latitudinal distribution of the vertical kinetic helicity $ h_k $ for depths 2.4 Mm (a) and 12.6\,Mm (c). There is a clear hemispheric preference in the parameter $h_k$~as there are $ 65\%(56\%)$ ARs at depth of 2.4\,Mm and  $69\%(67\%)$ ARs at depth of 12.6\,Mm in the northern (southern) hemisphere having negative(positive) values. The linear regression through ARs shows slopes of $-1.8\times10^{-10} s^{-2} degree^{-1}$  for the depth of 2.4\,Mm and $-1.2\times10^{-9} s^{-2} degree^{-1}$ for the depths of 12.6\,Mm. These results show that the kinetic helicity have same hemispheric preference as that of magnetic helicity $ \alpha_f $ and current helicity $ h_c $. The larger values of $ h_k $ for deeper layers indicate that the parameter $ h_k $ have better hemispheric preference at larger depths. 
		
		\begin{figure}[h]
			\centering \includegraphics[width=0.8\textwidth,clip,viewport=30 1 445 204]{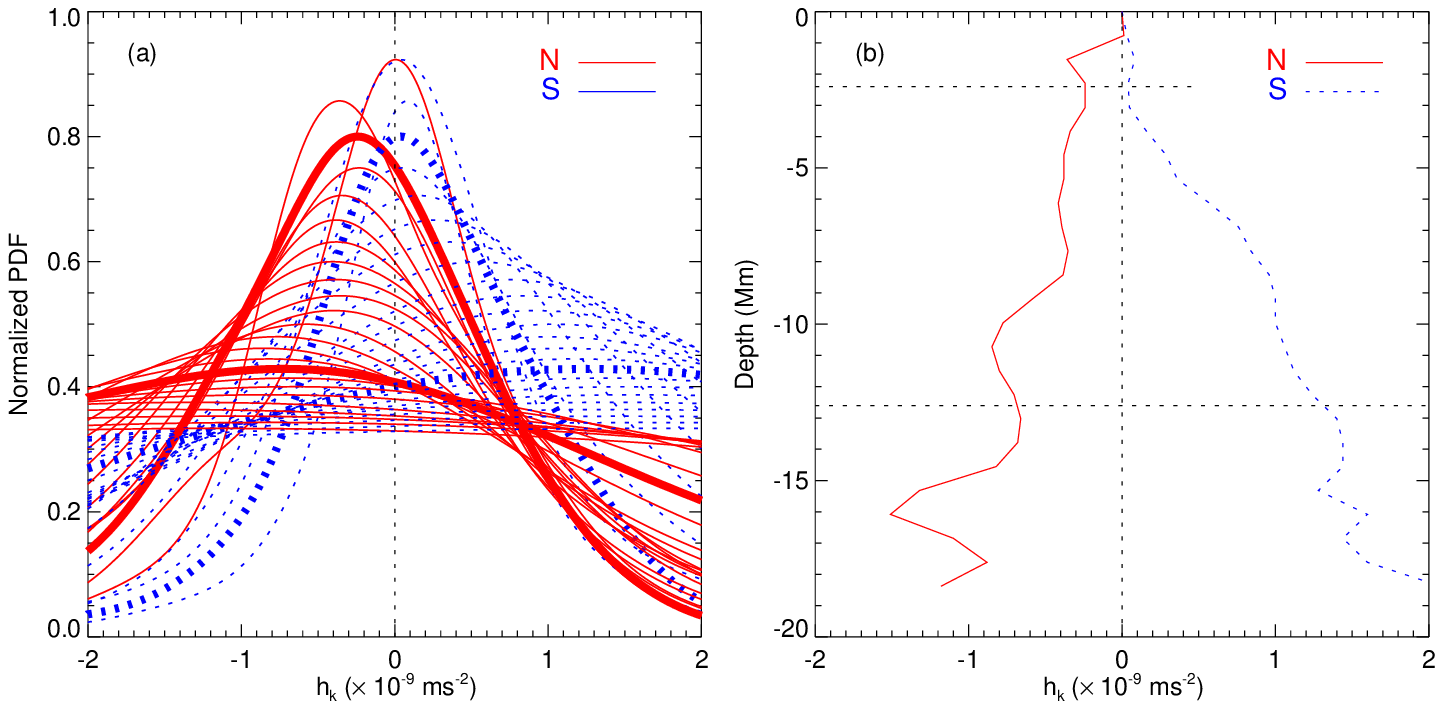}\\
			\centering \includegraphics[width=0.8\textwidth,clip,viewport=33 1 443 204]{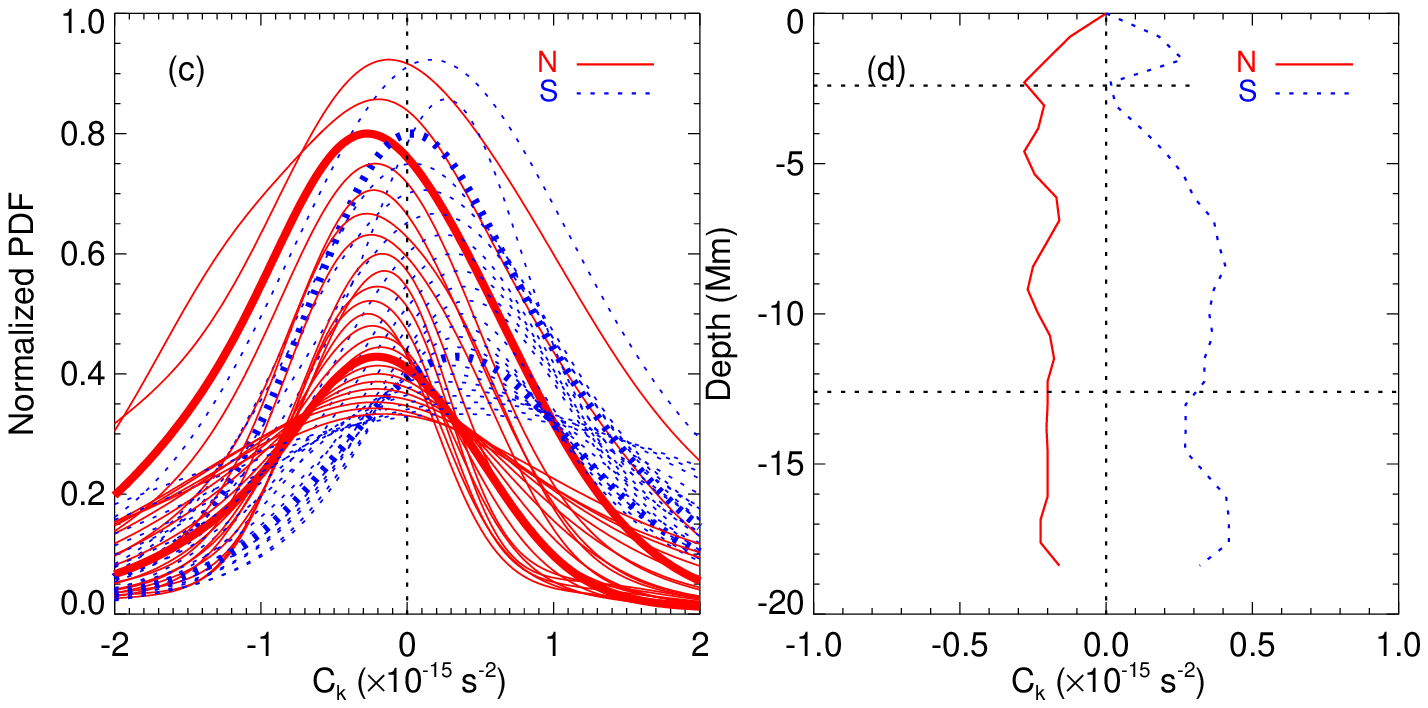}
			\caption{(a) Hemispheric probability density function for the sub-photospheric flow parameter $h_k$~for several depths. The PDF is scaled down for deeper layers. Thick curves correspond to depths of 2.4\,Mm and 12.6\,Mm (b) The peak point of the PDFs (in panel a) as a function of depth. Buttom row is similar to the top row but for the parameter $ C_k $.}
			\label{fig:kin-hel-depth-pdf}
		\end{figure}

		To further confirm the hemispheric preference of kinetic helicity $ h_k $, we have computed the PDF for the parameter $ h_k $ of northern and southern ARs. Figures~\ref{fig:kin-hel-lat-pdf}(b) and~\ref{fig:kin-hel-lat-pdf}(d) shows the PDF  of $h_k$ for the depths 2.4\,Mm and 12.6\,Mm. The PDF for $h_k$ peaked at $-0.2(+0.1)\times10^{-9} ms^{-2}$ for northern (southern) hemispheric ARs for the depth 2.4\,Mm. Similarly, the PDF for $h_k$ at the depth 12.6\,Mm peaked at $-0.7(+1.1)\times10^{-9} ms^{-2}$ which further confirms the HSR for parameter $h_k$.

		In order to check depth dependent HSR for $ h_k $, we computed the PDFs for all the available depths and plotted in Figure~\ref{fig:kin-hel-depth-pdf}(a). We have also estimated the $ h_k $ associated with peaks of PDFs and displayed in  Figure~\ref{fig:kin-hel-depth-pdf}(b). We find that the peak position shifted towards larger $ h_k $ values at larger depths for both hemispheric ARs.  The southern ARs show larger $ h_k $ values than northern ARs at deeper layers ($ >6$\,Mm) but at shallower depths ($ <6$\,Mm) northern ARs have larger $ h_k $ values than southern ARs. This could be due to stronger swirls in the sub-photospheric flows at deeper layers than shallower. 
		
		\begin{figure}
			\centering
			\includegraphics[width=0.45\textwidth,clip,viewport=19 6 274 272]{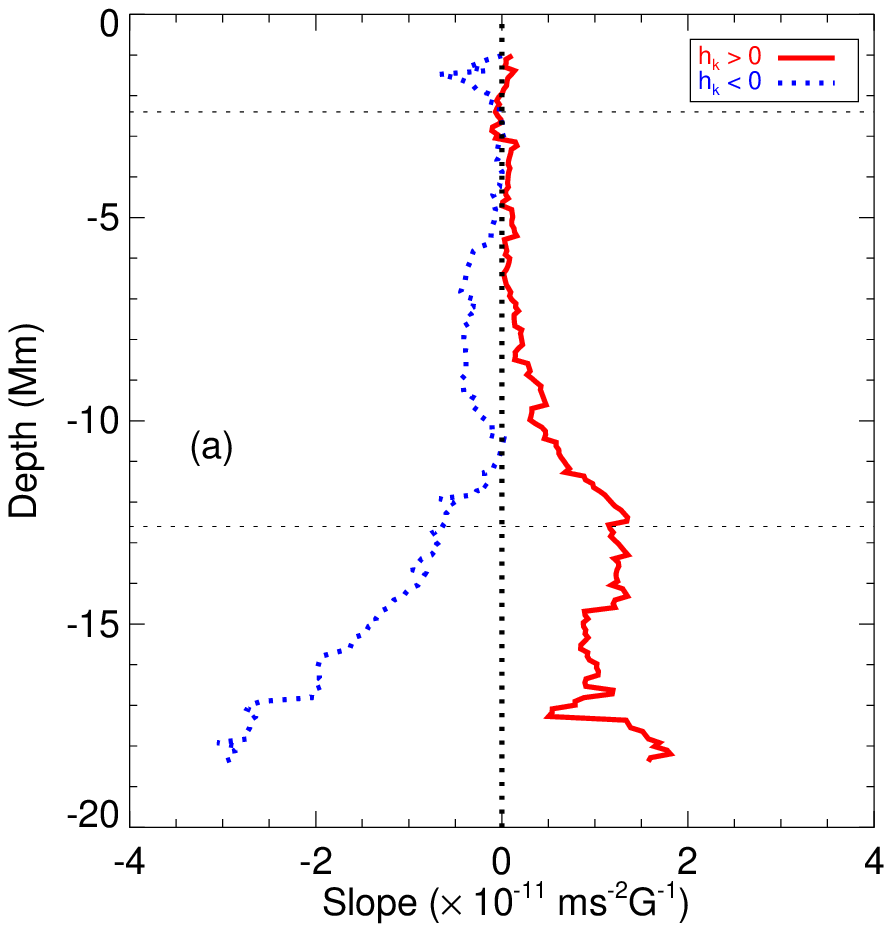}
			\includegraphics[width=0.45\textwidth,clip,viewport=19 6 274 272]{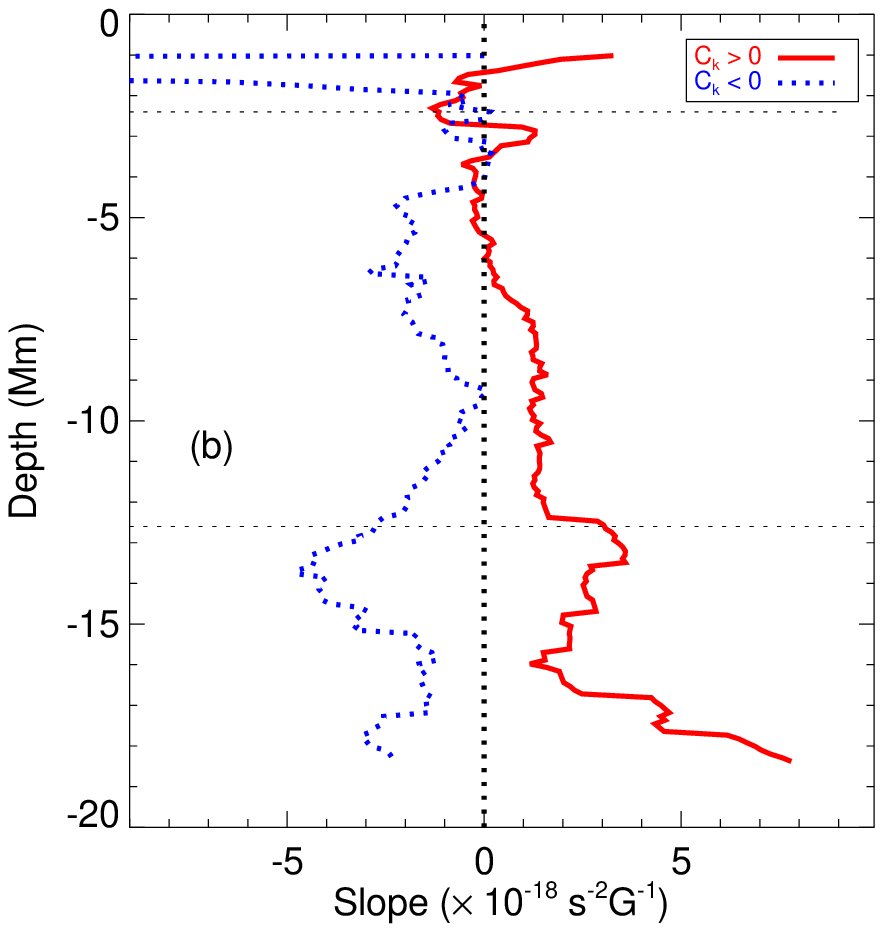}
			\caption{Variation in the slopes of the distribution of kinetic helicity $ h_k $ (a)  and divergence-curl $ C_k $(b) vs $ B $ as a function of depth for $ h_k>0 $ (solid and red) and $ h_k<0 $ (dotted and blue).}
			\label{fig:kin-hel-slope-b}
		\end{figure}

		We have also analysed the variations in the kinetic helicity as a function of magnetic flux $ B $. For this, we computed the slope of the distribution of $ h_k $ vs $ B $ as a function of depth (Figure~\ref{fig:kin-hel-slope-b}a). We found that the magnitude of the slope enhances considerably at larger depths $ >3\,Mm $ for both positive and negative helicities. For example, the slope for the depth 2.4\,Mm is found to be $ +1.9(+1.1)\times10^{-10}ms^{-2} $ for positive(negative) $ h_k $ values while it increased to $ +1.4(-1.3)\times10^{-9}ms^{-2}G^{-1} $ at the depth of 12.6\,Mm. The positive (negative) values of slope for the parameter $ h_k>0 $($ h_k<0 $) represent the linear relation between the helicity parameter $ h_k $ and $ B $. Similar result has been found for the magnetic helicity (Figure~\ref{fig:mag-hel-absB}b) and the current helicity (Figure~\ref{fig:mag-hel-absB}d).

		\begin{figure}
			\centering \includegraphics[width=1.0\textwidth,clip=,viewport=20 72 479 334]{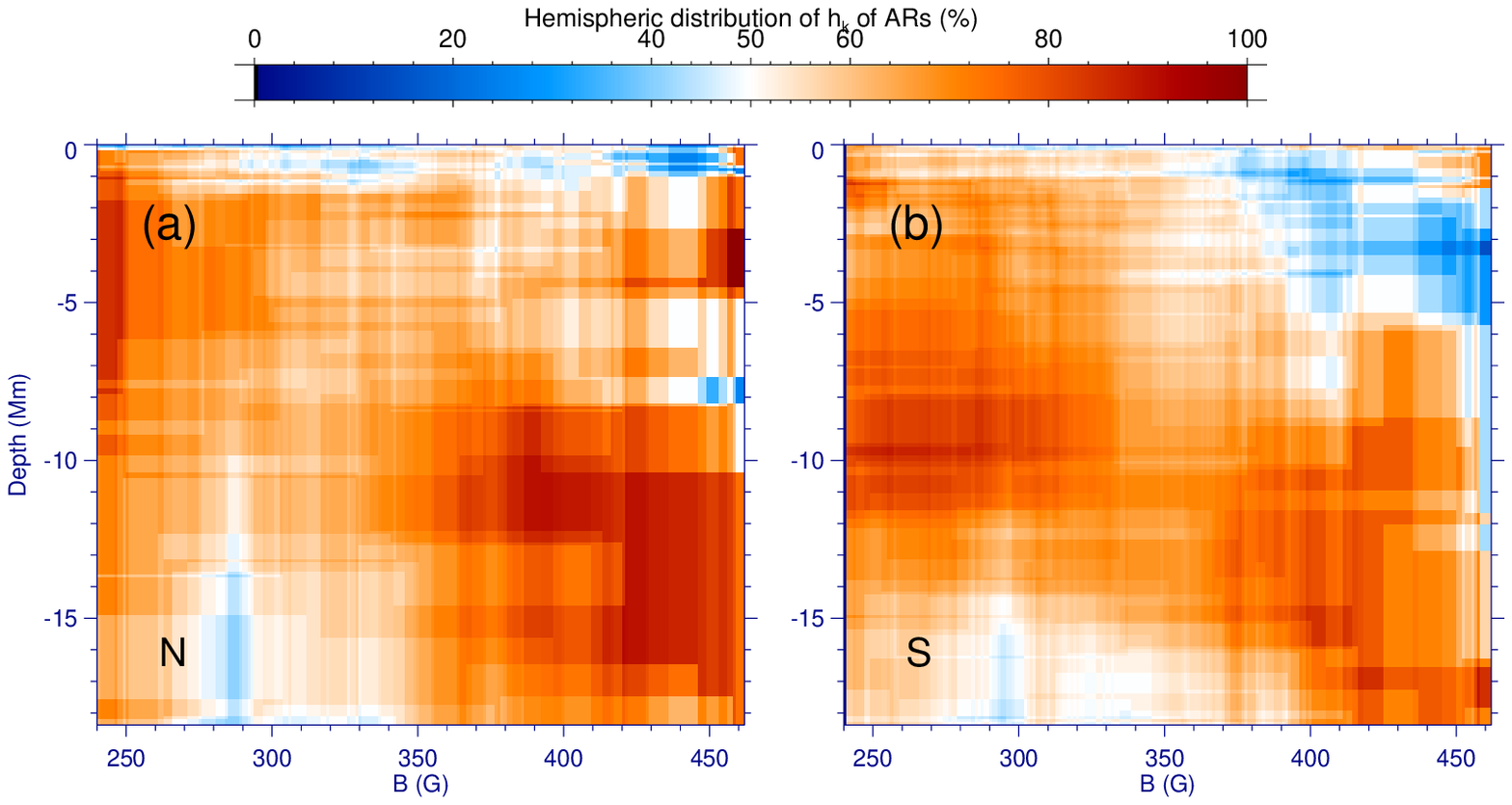}\\
			\includegraphics[width=1.0\textwidth,clip=,viewport=20 72 479 334]{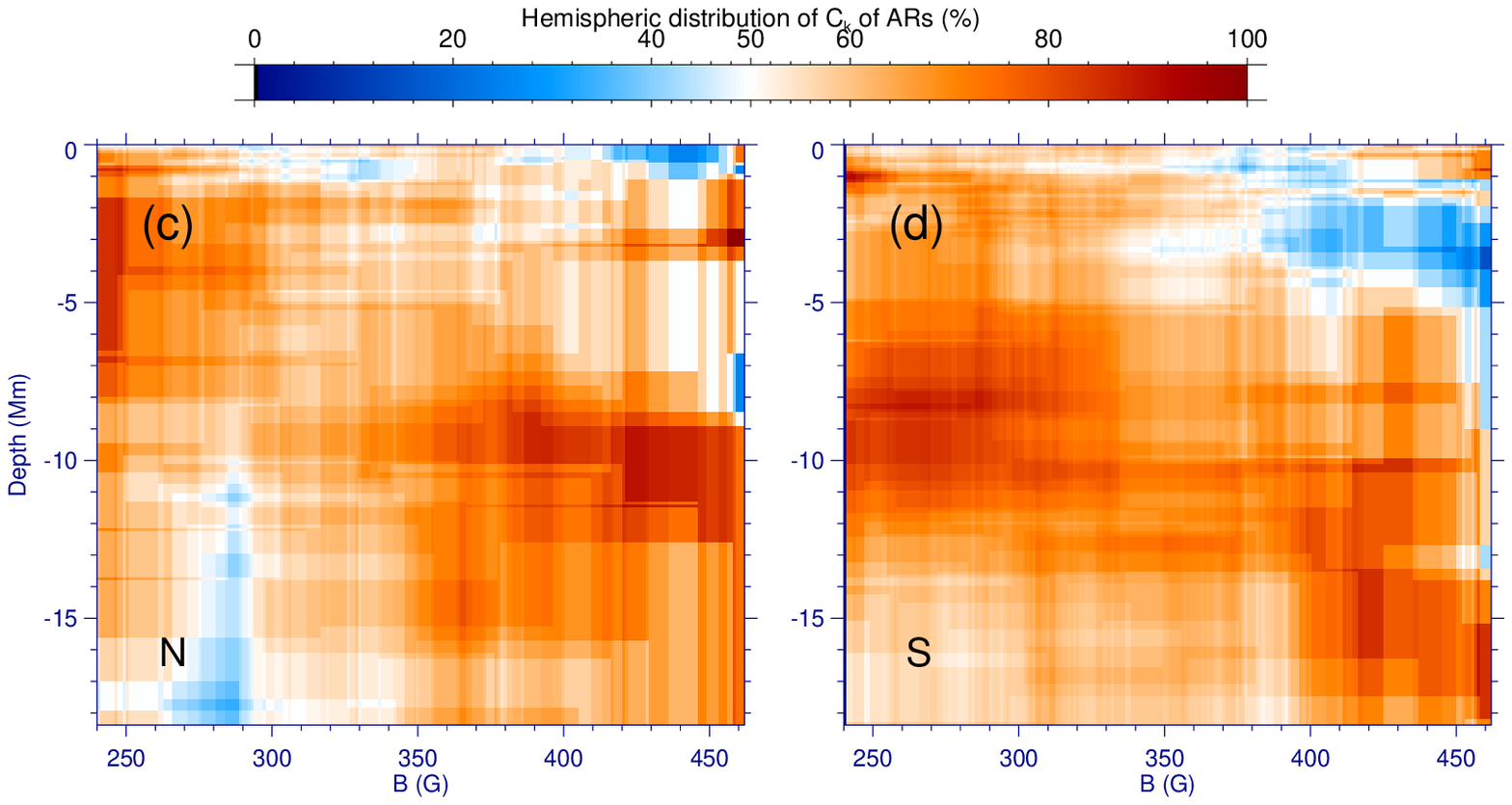}
			\caption{(Top row) Hemispheric distribution of sub-photospheric flow parameters $ h_k $ for northern (a) and southern (b) hemisphere, as a function of depth and magnetic flux $ B $. (Bottom row) Similar to the top row but for the parameter $ C_k $.}
			\label{fig:kin-hel-hemishp-map}
		\end{figure}		
	
		Next, we find the hemispheric distribution of the parameter $ h_k $ as a function of depth and magnetic flux. For this we counted the number of ARs following the HSR by running a magnetic flux window of width 50 G for all depths. These results are shown in Figure~\ref{fig:kin-hel-hemishp-map}(top row) for northern (left panel) and southern (right panel) hemispheres. This clearly shows that the HSR, statistically, becomes stronger at larger depths. Also, as the magnetic flux $ B $ increases the hemispheric biases of ARs shifted towards larger depths. However, in ARs with smaller fluxes show opposite HSR trend in both hemispheres. Similar preference is also discernible at shallower layers in ARs having stronger fluxes. The increase in hemispheric preference of with magnetic flux has also been found for the photospheric magnetic helicity $ \alpha_f $ (Figure~\ref{fig:mag-hel-absB}b) as well as for the current helicity $ h_c $ (Figure~\ref{fig:mag-hel-absB}d).

\begin{figure}
			\centering	
			\includegraphics[width=0.8\textwidth,clip,viewport=12 5 420 168]{sunspot_cyc_ars.eps}\\
			\includegraphics[width=0.8\textwidth,clip,viewport=14 5 420 241]{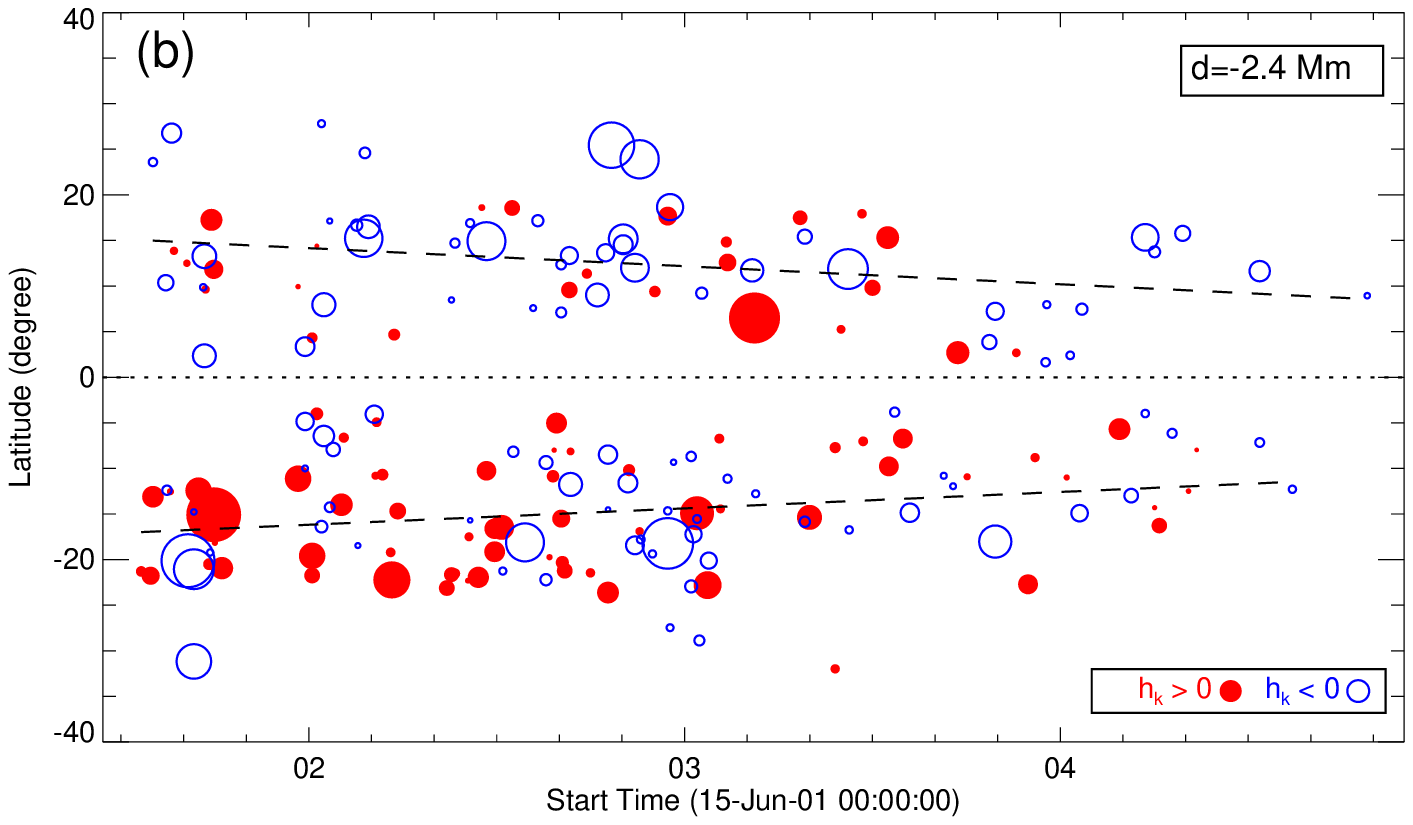}\\
			\includegraphics[width=0.8\textwidth,clip,viewport=12 5 420 241]{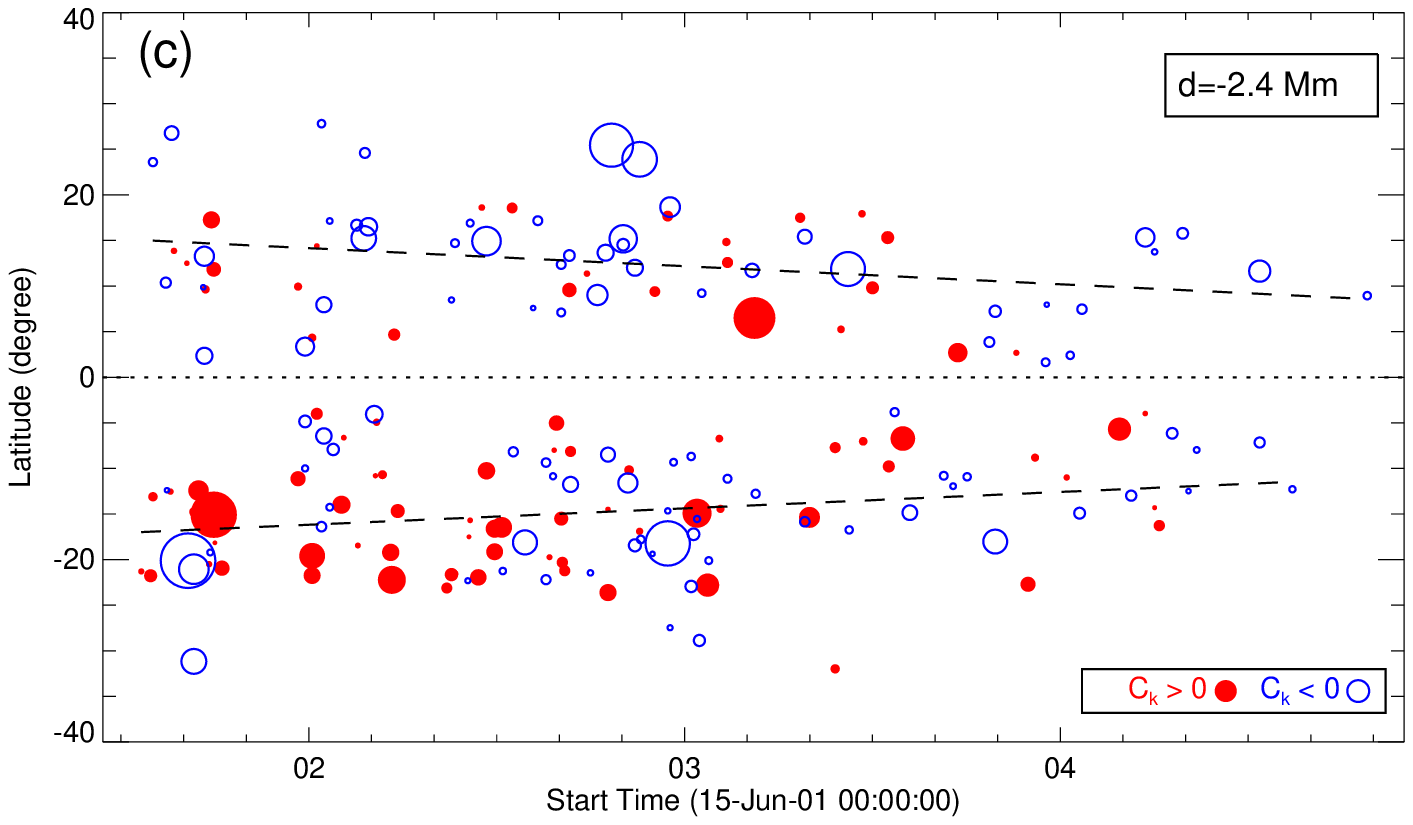}\\
			\includegraphics[width=0.8\textwidth,clip,viewport=17 7 420 165]{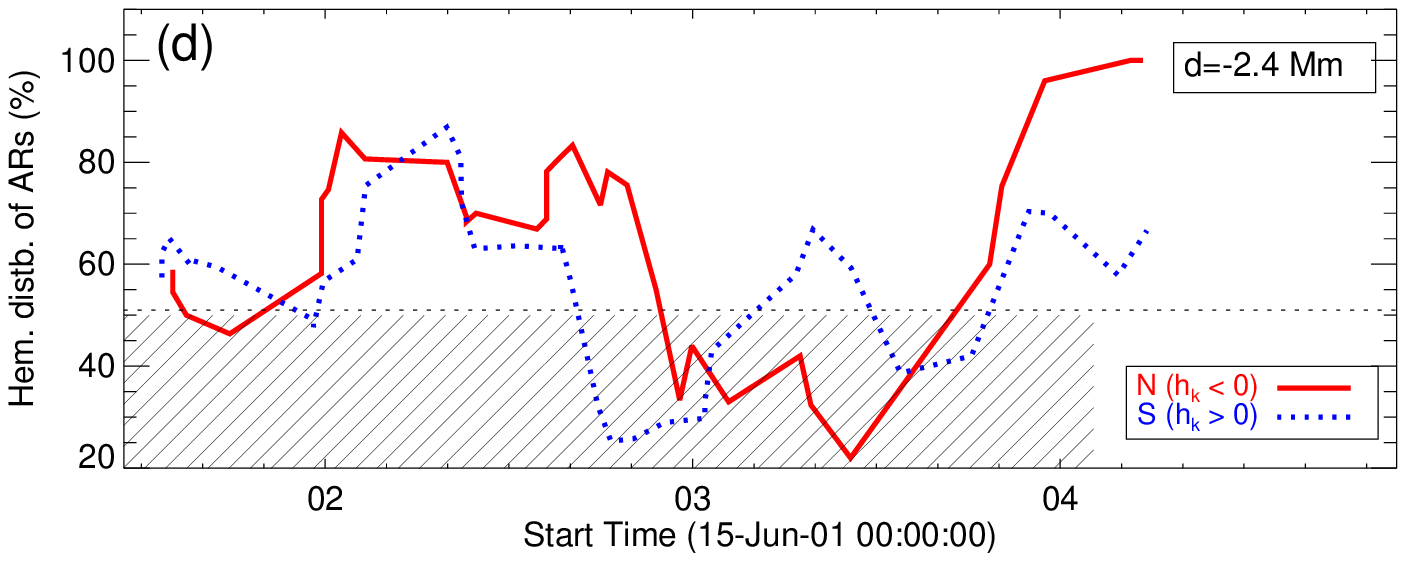}\\
			\caption{Similar to Figure~\ref{fig:mag-cur-hel-time} but for the parameters $h_k$ and $ C_k $ at depth 2.4\,Mm.}
			\label{fig:kin-hel-time-d1}
		\end{figure}

\begin{figure}
			\centering
			\includegraphics[width=0.8\textwidth,clip,viewport=12 5 420 168]{sunspot_cyc_ars.eps}\\ 
			\includegraphics[width=0.8\textwidth,clip,viewport=12 5 420 241]{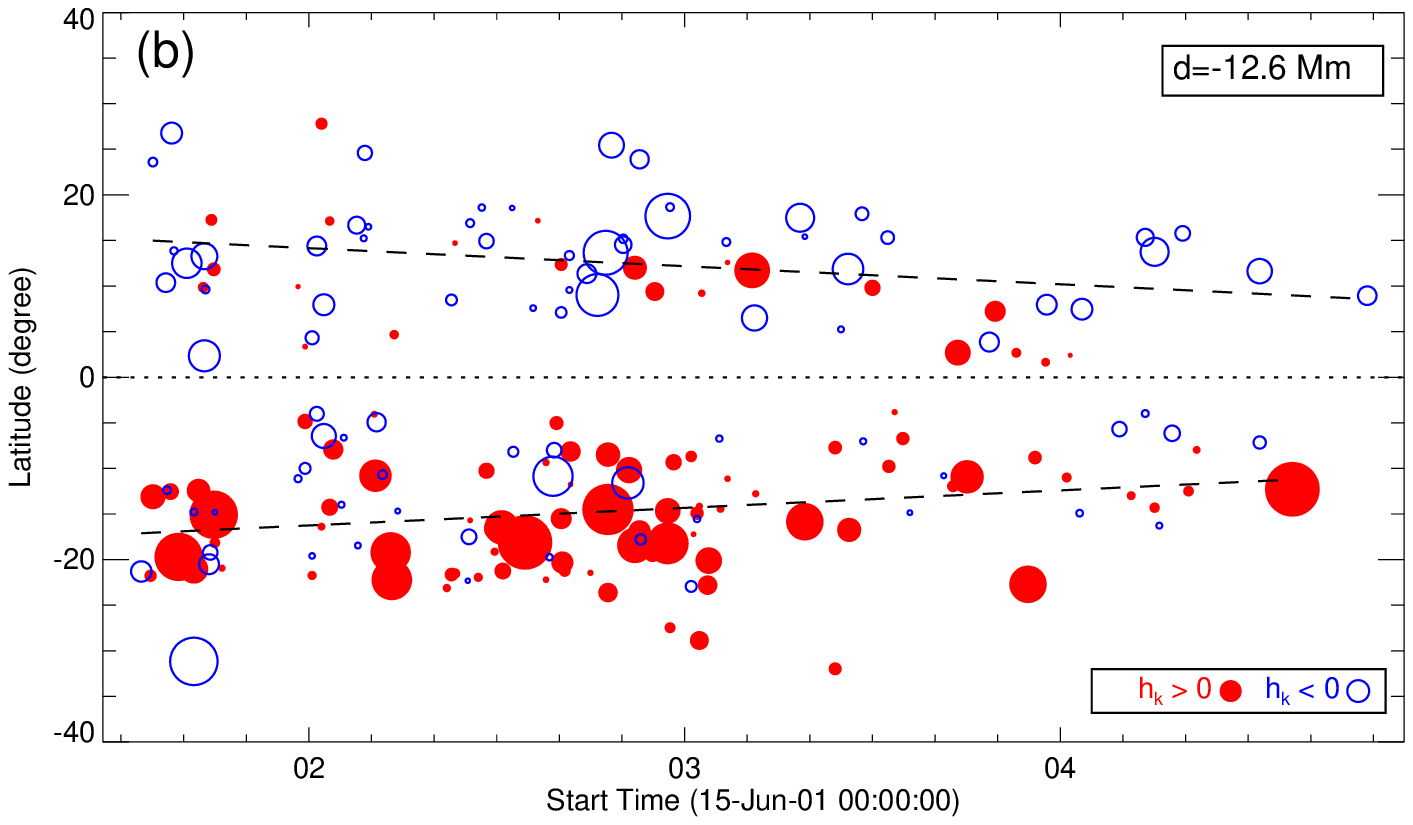}\\
			\includegraphics[width=0.8\textwidth,clip,viewport=12 5 420 241]{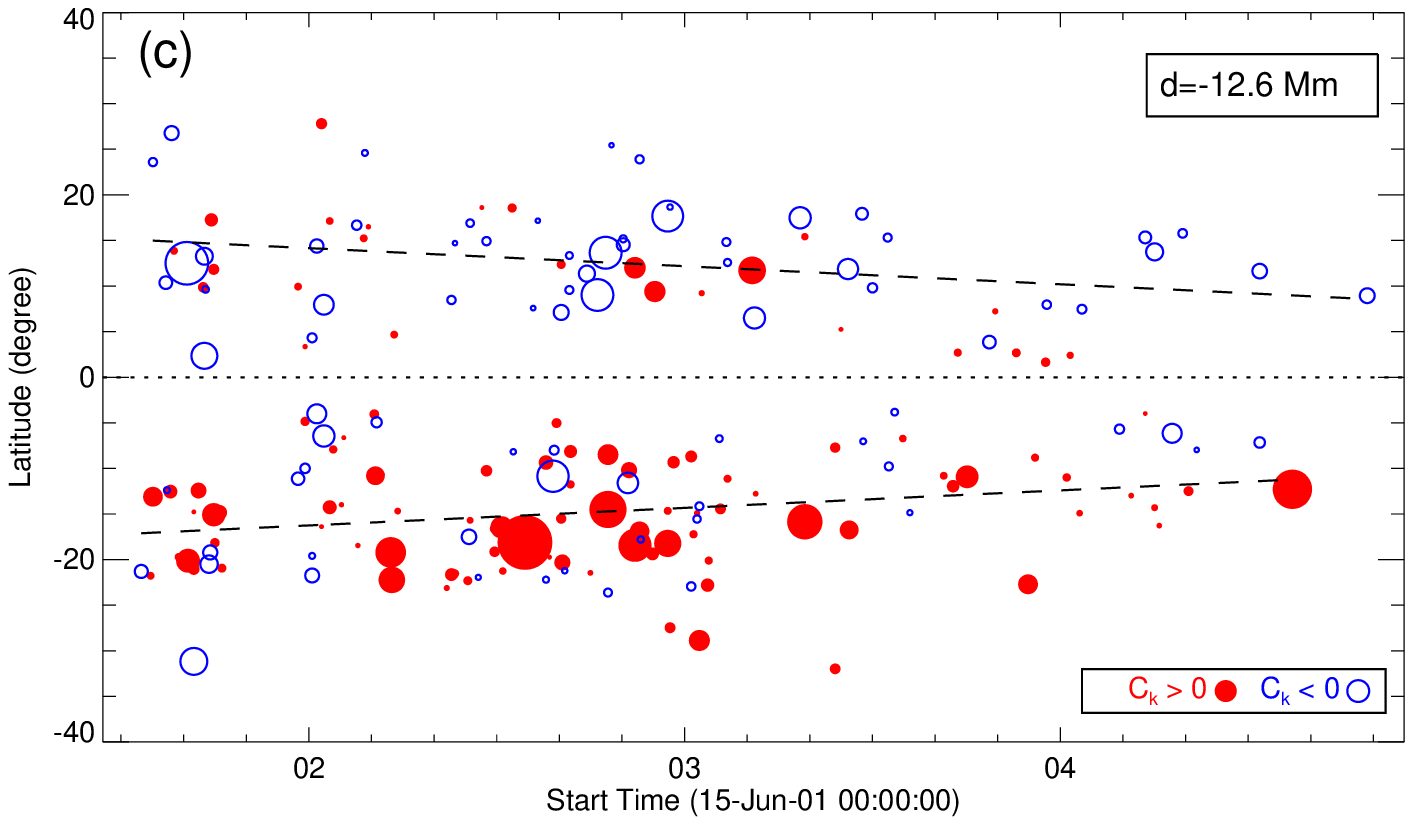}\\
			\includegraphics[width=0.8\textwidth,clip,viewport=17 7 420 165]{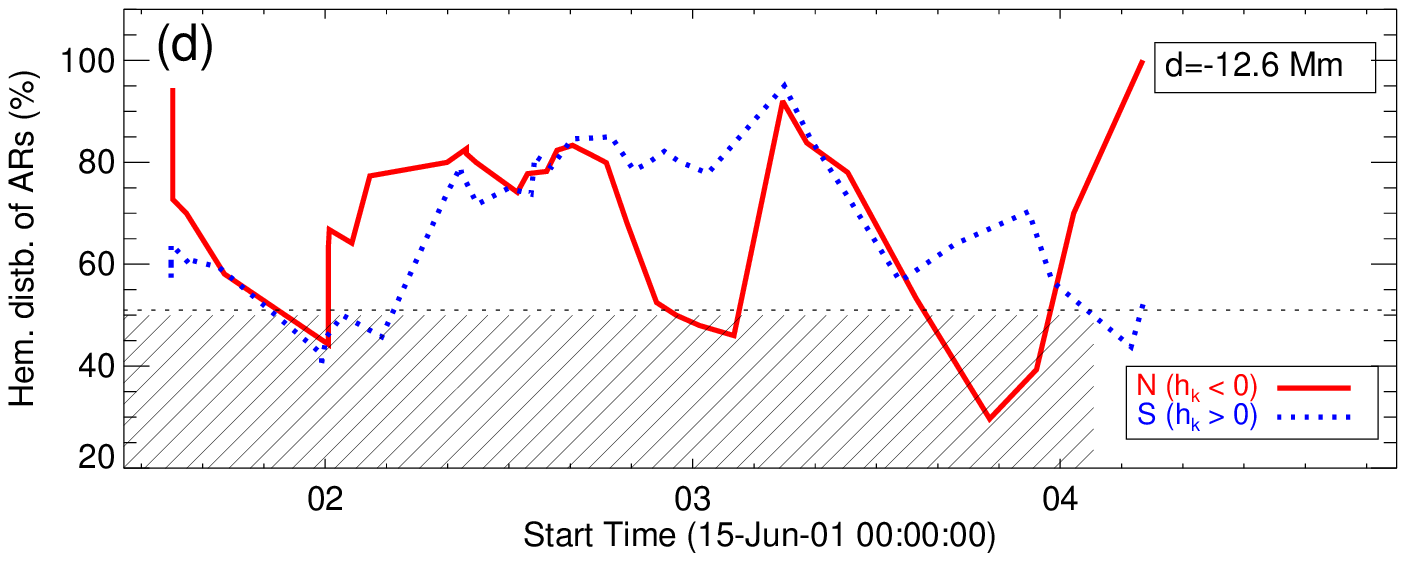}
			\caption{Similar to Figure~\ref{fig:kin-hel-time-d2} but for the depth 12.6\,Mm.}
			\label{fig:kin-hel-time-d2}	
		\end{figure}

		Figure~\ref{fig:kin-hel-time-d1}(b) and~\ref{fig:kin-hel-time-d2}(b) show the time dependent variations in the kinetic helicity parameter $ h_k $ for the depths 2.4\,Mm and 12.6\,Mm, respectively. The hemispheric biases are noticeable during the peak to decay phase of the solar cycle. But, these preferences are slightly weaker than the magnetic and current helicity parameters. The kinetic helicity $ h_k $ is found to be scattered significantly for both depths throughout the time period considered in the analysis. The magnitude of kinetic helicity of all ARs for the depth of 12.6\,Mm are found to be larger throughout than that for the depth of 2.4\,Mm. This shows largely variant sub-photospheric twisted flows in the interior of ARs. We could not find clear signature of equator-ward propagation of ARs for the parameter $ h_k $ at both depths.
		
		In the first half year of 2003, the ARs in  northern hemisphere deviated largely from usual HSR for the depth of 2.4\,Mm (Figure~\ref{fig:kin-hel-time-d1}d).  Similarly, southern ARs show opposite HSR during November 2002 to January 2003. For the depth of 12.6\,Mm (Figure~\ref{fig:kin-hel-time-d1}d), the ARs show better HSR even during the aforesaid period. Southern ARs show better HSR than northern ARs from the peak to decay phase of the solar cycle. But, at the end phase of the solar cycle, some ARs in southern hemisphere were found to have opposite hemispheric trend.
				
		\subsubsection{Divergence-Curls of Sub-photospheric Flows}
		\label{sbsbsec:kin-hel}
		
		Figure~\ref{fig:kin-hel-lat-pdf}(e-h) shows the latitudinal distribution of  sub-photospheric flow parameter $C_k$~for depths 2.4\,Mm (e) and 12.6\,Mm (g). For the above two depths, the latitudinal distributions of ARs are found to be $65\%(55\%)$ and $65\%(70\%)$ in the northern (southern) hemispheres having negative (positive) values of $C_k$. We noticed that the hemispheric distribution of ARs for $ C_k $ and $ h_k $ are same for the depth 2.4\,Mm but different for the depth of 12.6\,Mm. This is due to averaging of topological parameters of same ARs observed multiple times. The linear regression through distribution of the parameter $ C_k $ shows a slope of $ -2.4\times10^{-17} s^{-2} degree^{-1} $ for the depth of 2.4\,Mm and $ -2.5\times10^{-17} s^{-2} degree^{-1} $ for the depth of 12.6\,Mm. These results show hemispheric preferences similar to the magnetic helicity parameter $ \alpha_f $(Figure~\ref{fig:mag-cur-hel-lat}a), current helicity $ h_c $(Figure~\ref{fig:mag-cur-hel-lat}c) and the kinetic helicity parameter $ h_k $ (Figure~\ref{fig:kin-hel-lat-pdf}a and c).
		
		The probability density function for the sub-photospheric flow parameters $C_k$~is shown in Figures~\ref{fig:kin-hel-lat-pdf}(f) for the shallower depth at 2.4\,Mm and~\ref{fig:kin-hel-lat-pdf}(h) for the deeper layer at 12.6\,Mm. For the depth 2.4\,Mm the parameter $C_k$ peaks at $-2.5(+0.0\times10^{-16})\,{\rm s^{-2}}$ and for the depth 12.6\,Mm it peaks at  $-2.0(+3.6\times10^{-16})\,{\rm s^{-2}}$ for the northern (southern) hemisphere while average values of $ C_k $ were found to be $ -4.2(+0.22)\times10^{-16}s^{-2} $ for the depth of 2.4\,Mm and $ -0.43(+0.34)\times10^{-15} s^{-2}$ for the depth of 12.6\,Mm. This further confirms the HSR in $ C_k $ for the sub-photospheric flows of ARs. 
		
		Figure~\ref{fig:kin-hel-depth-pdf}(c) shows depth dependent PDF for the parameter $C_k$~for  northern (red) and southern (blue) hemispheres similar to Figure~\ref{fig:kin-hel-depth-pdf} (a) for the kinetic helicity $ h_k $. Figure~\ref{fig:kin-hel-depth-pdf}(d) represents the variations in peak position of PDF for northern (red) and southern(blue) hemispheres as function of depth. The hemispheric distribution for the parameter $C_k$~is found to exist for all depth range 0 -- 20\,Mm. At depths beyond 2.0\,Mm it is significant as evident from Figure~\ref{fig:kin-hel-depth-pdf}(d).  The peaks of PDFs, between depth range 5 -- 20\,Mm, for both hemispheres are found to be nearly constant at $ -2.0(+3.5)\times10^{-16} s^{-2} $ for northern (southern) hemisphere while it shifted towards larger values at larger depth for the parameter $ h_k $. The constant distribution of maximum peak of PDF of $ C_k $ is due to horizontal divergence. The flows are found to be largely diverging near the photosphere than in deeper layers. Hence increase in vorticity is compensated by decrease in divergence and results into constant value for the parameter $ C_k $.
		
		In order to study the variation in the topological parameter $ C_k $ as a function of depth and magnetic fields, we computed the slope of the distribution, $ C_k $ vs. $ B $. The results are shown in the Figure~\ref{fig:kin-hel-slope-b}(b) for $ C_k>0 $(solid and red curves) and $ C_k<0 $(dotted and blue curves). For example, the slope for $ C_k>0 $ ($ C_k<0 $) for the depth of 2.4\,Mm is found to be $ +0.78(-3.04)\times10^{-18}s^{-2}G^{-1} $ while it changed  to $ +1.61(-2.63)\times10^{-18}s^{-2}G^{-1} $ for the depth of 12.6\,Mm. The slopes, on average, are found to increase as we go deeper ($ >4\,Mm $).

		Similar to the helicity $ h_k $ (Figure~\ref{fig:kin-hel-hemishp-map}, top row), we computed the depth and magnetic field dependent hemispheric preference of the parameter $ C_k $. This result is shown in Figure~\ref{fig:kin-hel-hemishp-map}(bottom row). It clearly shows that the hemispheric distribution of $ C_k $ shifted towards larger depths as the magnetic field increases. Also, it can be noticed that the HSR became stronger with depth and magnetic flux $ B $. Near the photosphere with stronger flux regions and deeper layers with smaller fluxes, the opposite hemispheric trends have been found in both the hemisphere.  The hemispheric preference for the parameter $ C_k $ is very similar to that of $ h_k $ (Figure~\ref{fig:kin-hel-hemishp-map}, top row) with same sign of handedness.

		In order to check the solar cyclic variation of the parameter divergence-curl ($ C_k $), we have plotted it as a function of time for two depths 2.4\,Mm (Figure~\ref{fig:kin-hel-time-d1}c) and 12.6\,Mm (Figure~\ref{fig:kin-hel-time-d1}c). We found that most of ARs shows $ C_k $ around an absolute mean of $ 1.0\times10^{-15} s^{-1} $ for both the depths. We found that the parameter $ C_k $ ranges between $ \pm4\times10^{-15}s^{-2} $ for both the depths (Figure~\ref{fig:kin-hel-time-d1}c and~\ref{fig:kin-hel-time-d1}c) while $ h_k $ (Figure~\ref{fig:kin-hel-time-d1}b and~\ref{fig:kin-hel-time-d1}b) have larger ranges for deeper depths. Similar to the parameter $ h_k $, we did not find clear evidence of equator-ward propagation of the parameter $ C_k $.

		\subsection{Relation between Kinetic and Magnetic Helicities}
		\label{sbsec:rel-kine-mag-hel}
		
\begin{figure}[H]
			\centering	
			\includegraphics[width=0.49\textwidth,clip,viewport=38 1 512 203]{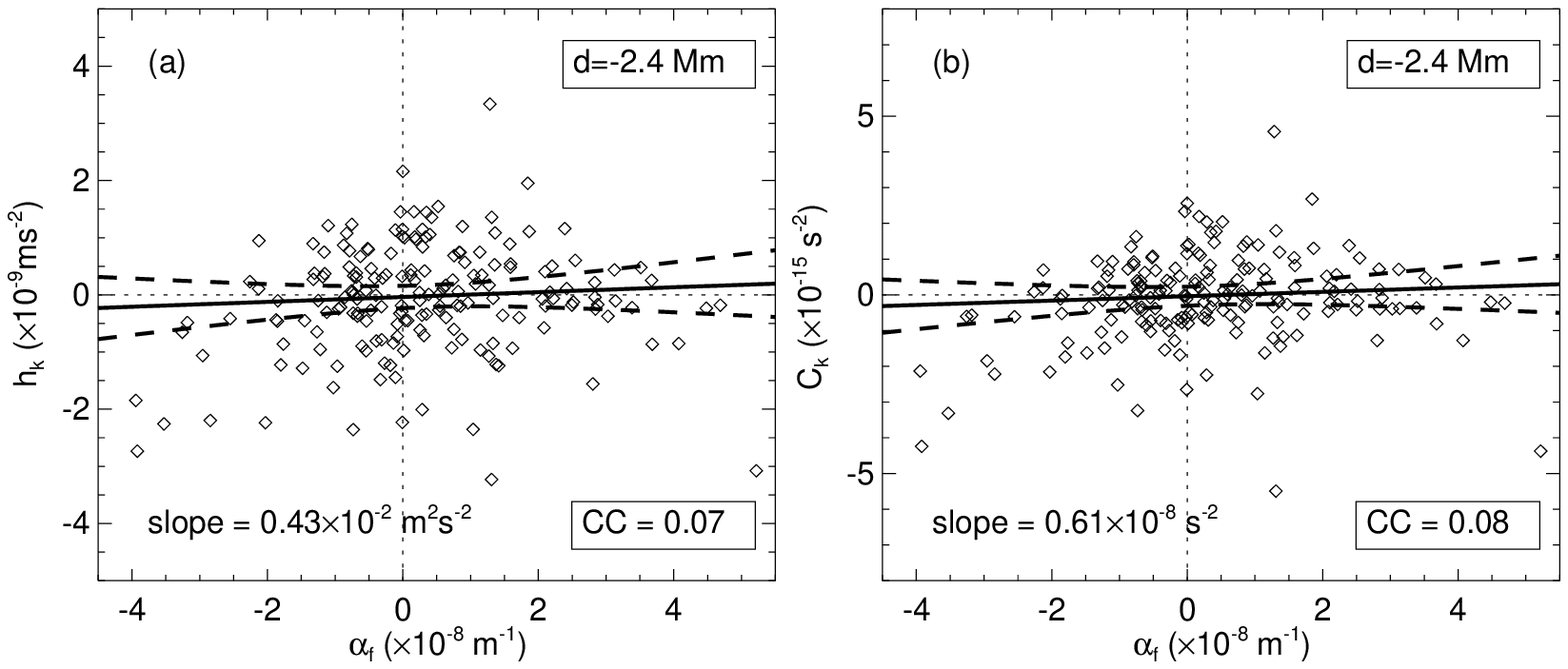}
			\includegraphics[width=0.49\textwidth,clip,viewport=38 1 512 203]{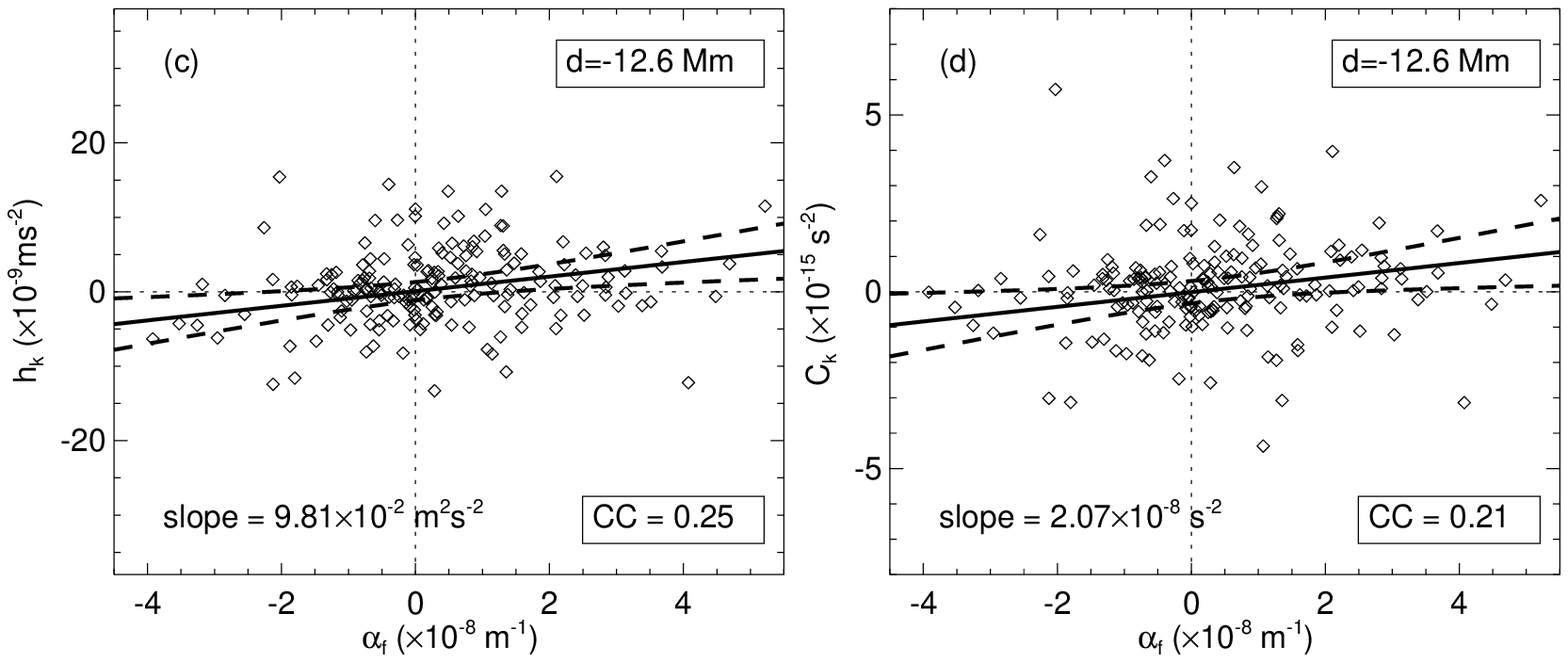}
			\includegraphics[width=0.49\textwidth,clip,viewport=38 1 512 203]{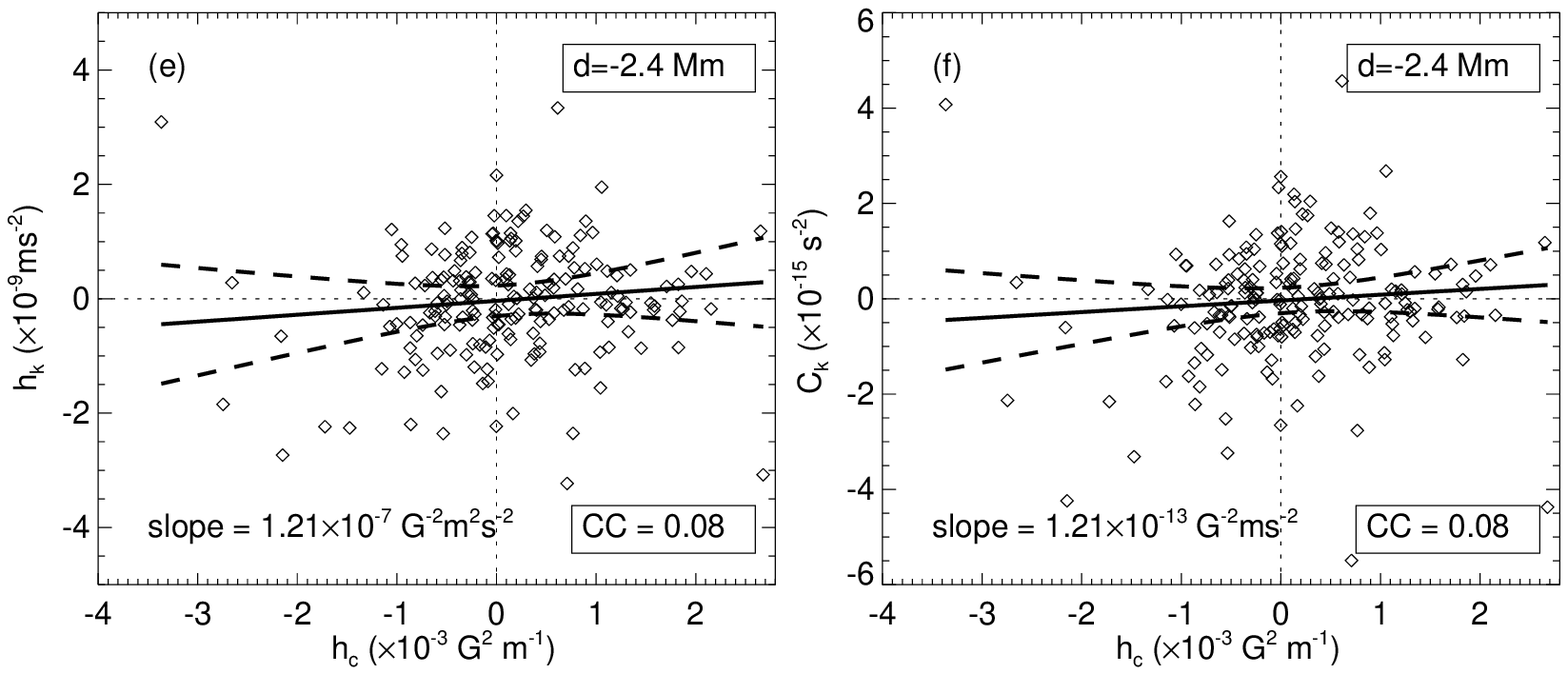}
			\includegraphics[width=0.49\textwidth,clip,viewport=38 1 512 203]{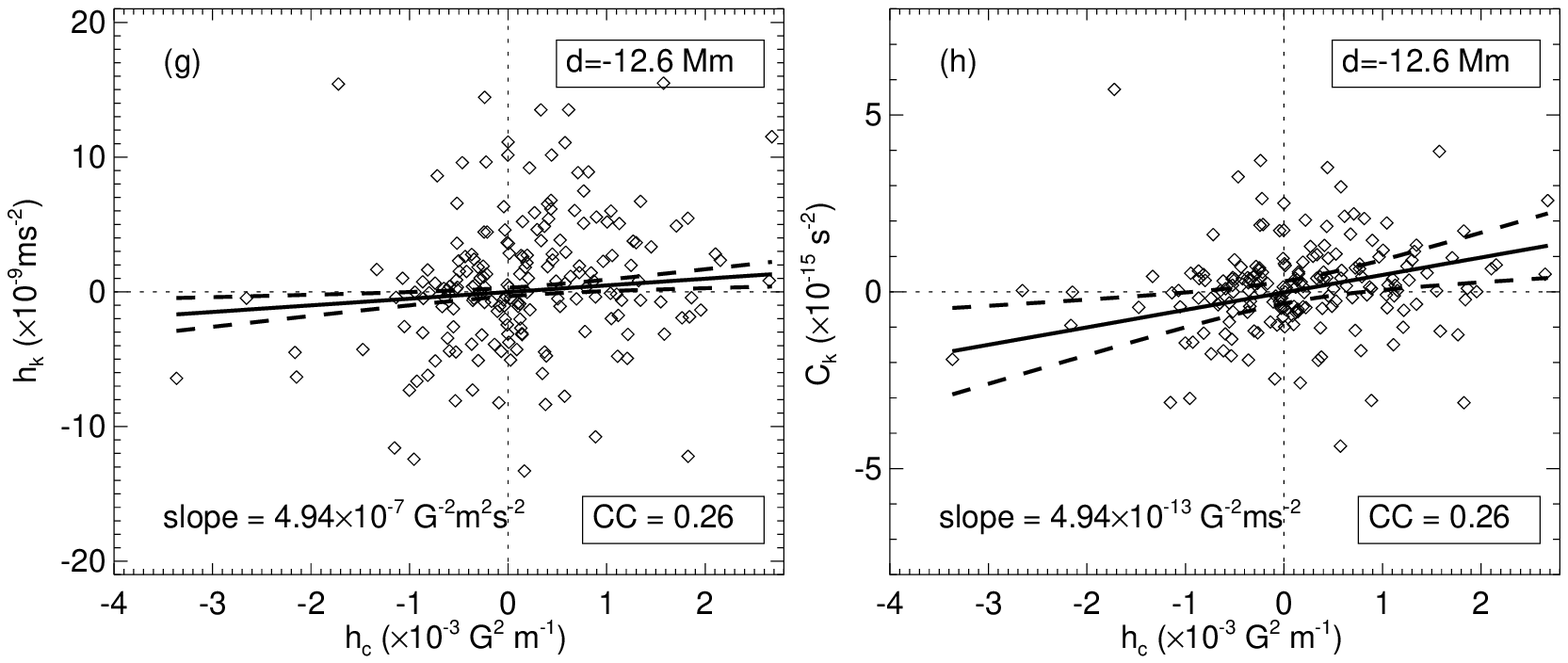}
			\caption{(top row) Kinetic helicity parameters $ h_k $ and $ C_k $ for two  depths 2.4\,Mm (a and b) and 12.6\,Mm (c and d), as a function of magnetic helicity parameter $ \alpha_f $, where the values of $ CC $ represent the Pearson correlation coefficients between the topology of photospheric magnetic fields and sub-photospheric flows. Bottom row is similar to the top row but as a function of current helicity $ h_c $.}
			\label{fig:assoc-kin-mag-hel}
		\end{figure}

\begin{figure}
			\centering	
			\includegraphics[width=1.0\textwidth,clip,viewport=40 7 475 235]{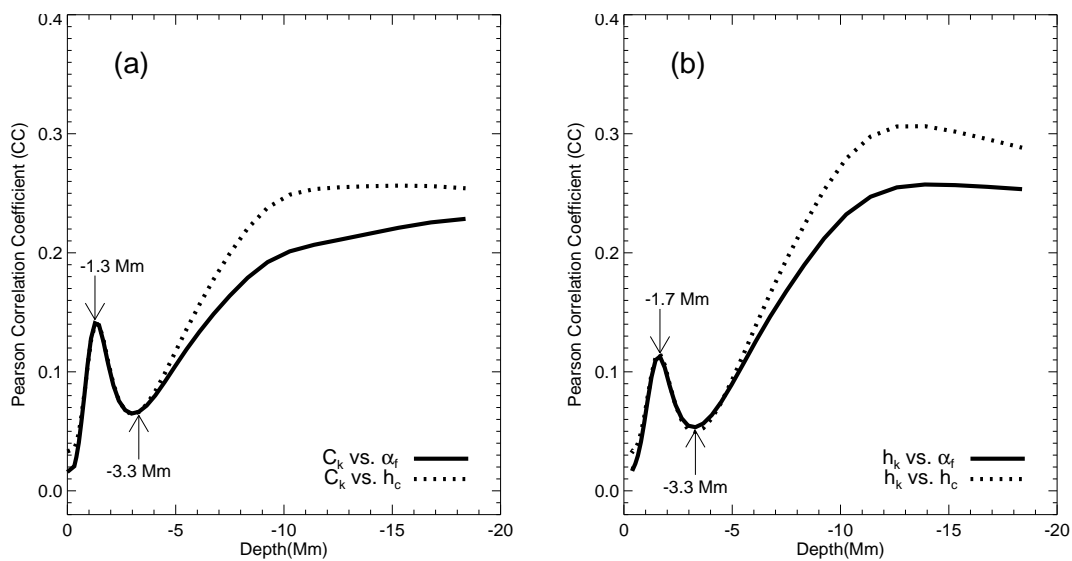}
			\caption{Depth dependent correlation coefficients between kinetic helicity and magnetic helicity parameters: (a) $ C_k $ \& $ \alpha_f $ (solid line), $ C_k $ \&  $h_c $ (dashed line), (b) $ h_k $ \& $ \alpha_f $ (solid line), $ h_k $ \&  $h_c $ (dashed line).}
			\label{fig:correl-kin-mag-hel-depth}
		\end{figure}

		The relation between the surface and sub-surface helicities are shown in  \mbox{Figures~\ref{fig:assoc-kin-mag-hel} --~\ref{fig:correl-kin-mag-hel-depth}}. Figure~\ref{fig:assoc-kin-mag-hel}(top row) shows the correlation between the force-free parameter $\alpha_f$~and the sub-photospheric twist parameters for the parameters $h_k$ (a and c) and $ C_k $ (b and d) for two depths 2.4\,Mm (a and b) and 12.6\,Mm (c and d). The parameter $ \alpha_f $ shows correlations with $ C_k $ and $ h_k $ at both depths. The correlation coefficient is significantly larger for deeper layer than the shallower. The significant values of the correlation coefficient $ CC $ shows relation between the two topological parameters of the surface magnetic fields and the sub-photospheric flows.

		Figure~\ref{fig:assoc-kin-mag-hel}(bottom row) shows correlation between the current helicity and the twist parameter of sub-photospheric flows similar to Figure~\ref{fig:assoc-kin-mag-hel}(top row) for the parameter $ \alpha_f $ and sub-photospheric flow parameters $ h_k $ and $ C_k $. Corresponding correlation coefficients are given in respective panels. The magnitudes of correlations imply  association between the current helicity and the twist of sub-photospheric flows of ARs.
		
		In order to check the depth dependent association between topology of photospheric magnetic fields and sub-photospheric flows, we have computed the correlation coefficients between magnetic and flows helicities parameters as a function of depth. The result is shown in  Figure~\ref{fig:correl-kin-mag-hel-depth}. 
		
		An interesting pattern of $ CC $ with depth is that it is not random. It changes gradually with depths. Very near the photosphere the sub-photospheric flows and magnetic fields topology show no significant relation as the $CC$ is close to zero. But the absolute $ CC $ gradually increases with depth up to around of 1.5\,Mm, then decreases and becomes minimum around a depth of 3.3\,Mm, and thereafter starts increasing. There is a sharp increase in $ CC $ between 3.4 -- 12\,Mm and after that it becomes almost constant.  This shows that the photospheric magnetic field lines and the sub-photospheric flows at larger depths have similar topological trend. The two peaks in the $ CC $-profile indicates the bi-polar structure in the sub-photospheric flows as reported earlier \citep{Komm2005a,Maurya2009b,Maurya2010d}.
		
		Another, interesting pattern of $ CC $ is that at larger depths (> 5\,Mm) the magnitude of $ CC $ between current helicity and sub-photospheric topology is larger than the magnitude of $ CC $ between force-free parameter $ \alpha_f $ and sub-photospheric topology parameters. This shows that the current helicity has association with the sub-photospheric flows topology. This is also evident from the mean field dynamo \citep{Biskamp2003}.
		
		The gradual variation of $ CC $ between surface and sub-surface helicities could be due to the relation between the magnetic field and sub-photospheric flows. But, one should note that we are comparing the topology of magnetic field at the photosphere and the flows at depths. Since, there is no direct measurement of magnetic field in sub-photospheric depths, we cannot confidently suggest their association with flows. But from the above results it suggests that they may have some relation.

		\section{Summary and Conclusions}
		\label{sec:sum-conc}
		
		Using MSFC vector magnetograms and GONG Dopplergrams observations, we have analyzed the topological properties of photospheric magnetic fields and sub-photospheric flow fields of 189 ARs that cover  peak to descending phase of the solar cycle 23. Our analysis confirms previous reports, and show some new interesting results.
		
		We found clear evidence of hemispheric preferences for the magnetic and current helicities. The ARs of the northern(southern) hemisphere are found with dominantly negative (positive) helicities which confirms earlier reports \citep[][and others]{Pevtsov2008}. However, many ARs were found to show opposite hemispheric preferences, especially at beginning and end phase of the solar cycle as indicated in others studies \citep{Bao2000,Tiwari2009a, Pipin2019}. One should note that in complex magnetic field regions magnetic helicity proxy $ \alpha_f $ may deviate from the force-free field, \ie, $ \alpha_f $ may not be a true proxy for helicity in complex sunspot regions to correctly reflect the twistedness of an AR as conjectured by \citet{Russell2019}.  
		
		We have also found signature of equator-ward propagating pattern in the magnetic and current helicities similar to sunspots, in both hemispheres, during descending phase of the solar cycle. \citet{Zhang2010} have also reported similar results and suggested that the solar dynamo could be helical in nature. An AR with larger area and stronger fields may have larger helicities. Therefore such patterns of helicities could also be due to variations of areas and magnetic fields of sunspots during a solar cycle, from larger areas with stronger fields regions to smaller areas with weaker field regions \citep{Watson2011}. 
		
		On comparing the magnetic and current helicity parameters with field strength of ARs, we found that the magnitude of both helicities increases. Another interesting pattern was found that the hemispheric preferences become stronger for both the helicity types, implying that ARs with larger field strengths show better hemispheric preferences. Hemispheric trend of current helicity infers  that the regular vortex structure of sunspots is caused by forces such as the Coriolis and differential rotations.
		
		In the sub-photospheric flows, we found similar hemispheric trend for the topological parameter, kinetic helicity  $ h_k $ and divergence-curl $ C_k $ at all depths. Both the flow parameters follow same hemispheric preference as that of magnetic and current helicity parameters, \ie, $ h_k $ and $ C_k $ are found to have dominantly negative (positive) sign in ARs of northern(southern) hemisphere. This confirms the earlier reports \citep{Maurya2011b, Seligman2014, Komm2014a, Komm2015b}. But \citet{Gao2009} have found opposite hemispheric preferences in a small sample of ARs. Further, we could not detect clear signature of equator-ward propagating patterns in kinetic helicities of ARs as found for the magnetic helicities. Since kinetic and current helicities are related by mean field dynamo model such patterns are expected to exist in both the parameters.
		
		The hemispheric preference of the parameter $ h_k $ could be due to mean field dynamo $\alpha_d$ \citep[][see Equation 5.19]{Biskamp2000}. Let us think of the hemispheric dependence of force-free alpha $ \alpha_f $, dynamo alpha $ \alpha_d $ and kinetic helicity, $ h_k=<\mathbf{ u}\cdot\nabla\times\mathbf{ u}>\approx3<u_z(\nabla\times\mathbf{ u})_z> $. According to the mean-field dynamo, $ \alpha_d $ and $ h_k $ are strongly related to each other by the relation $ \alpha_d=-\frac{1}{3}\tau<\mathbf{u}\cdot\nabla\times\mathbf{u}>=-\frac{1}{3}h_k\tau $, where $\tau$ is the velocity correlation time. Thus $ h_k $ and $ \alpha_d $ have opposite signs. In the northern hemisphere, the Coriolis force causes a flow to be left-helical and $ h_k<0 $. From our analysis, we have found statistically negative kinetic helicity for ARs which support the mean-field dynamo theory. Further, considering $ h_k<0 $, we come to have $ \alpha_d >0$ in the northern hemisphere. This is commonly adopted view in the solar dynamo. It also seems to be consistent with Joy's law \citep{Hale1919}. The condition of $ \alpha_d>0 $ as well as the Joy's law suggests that the larger-scale magnetic field comprising bipolar ARs may be right helical. On the other hand, the force-free alpha $ \alpha_f $ is empirically found to be negative in the northern hemisphere. This means that magnetic fields are left-helical in scales smaller than individual ARs.
		
		The solar differential rotation may be another factor for the hemispheric trend of the sub-photospheric flows as rotation may introduce left- and right-handed vorticity in northern and southern hemispheres with magnitude of $0.01\times10^{-6} s^{-1}$ to $2.5\times 10^{-6} s^{-1}$ for an AR of sizes 1\,Mm to 40\,Mm along the meridional direction.
		
		The hemispheric preference for kinetic helicity is found to increase with depths. The magnitude of peak of kinetic helicity distribution is found to increases with depth for both hemispheric ARs while divergence-curl is found to be constant around depths 5 -- 20\,Mm. The reason for growth in kinetic helicity with depth is mainly due to increase in vertical velocity while constant value of the parameter $ C_k $ is caused by net effects of increase in swirl and decrease in divergence. 
		
		Our analysis shows interesting patterns of hemispheric preferences, of sub-photospheric flow topology, as a function of magnetic field strength and depths. The HSR for sub-photospheric flows topology enhances with magnetic field strength and with depths. Poor hemispheric preferences of flow topology near the photosphere, may be due to smaller plasma density and diverging magnetic field lines. This may be due to different scales of magnetic field lines. This supports the solar dynamo theory which predicts the bi-helical properties of magnetic fields \citep{Blackman2002,Brandenburg2005}. 
		
		Further, we analysed the correlation between sub-photospheric flows and photospheric magnetic fields. The kinetic helicity, near the photosphere, do not show significant relation with magnetic or current helicities. But we found that the correlation between them enhances as we go deeper into the solar interior. The kinetic helicity and vertical divergence-curl show correlation with magnetic and current helicities at all depths. Further, we found that the kinetic helicity shows better covariance with magnetic and current helicities than divergence-curl of flow fields, which supports the mean field dynamo models. Near-surface flow topology shows poor correlation with magnetic and current helicities. This suggests that the magnetic fields and flows may have different topologies. Another reason could be the frozen-in field conditions in the interior plasma, which may not be satisfied at these depths and results into weak association of magnetic and flow fields. But deeper into the interior there is increasingly high density plasma. There the field lines may be concentrated close to the flux tubes which may lead to similar topology in both physical quantities.

		It should be noted that we have used a single (or a few) vector magnetograms map corresponding to an AR to derive the photospheric twist while the twist in sub-photospheric flow was obtained using the data cube for 1664 minutes. This could be one reason for weak relation between the photospheric magnetic fields and sub-photospheric flows. But magnetic helicity is likely to be approximately conserved in solar atmosphere on time scale of AR while passing the disc \citep{Berger1984a}. It should also be noted that we have used magnetic and Doppler fields observations obtained from different sources. These factors may be responsible in affecting our results and in reducing correlations of the twist parameters. It is expected that these results may improve in the future with magnetic and Doppler velocity observations obtained from the same instrument.

		\acknowledgements{This work utilizes data obtained by the GONG program operated by AURA, Inc. and managed by the National Solar Observatory under a cooperative agreement with the National Science Foundation, U.S.A. The vector magnetogram data were obtained from MSFC. This work also utilizes data from the Solar Oscillations Investigation/Michelson Doppler Imager (SOI/MDI) on the Solar and Heliospheric Observatory (SOHO). SOHO is a project of international cooperation between ESA and NASA. MDI was supported by NASA grants NAG5-8878 and NAG5-10483 to Stanford University. We are grateful to Prof. J. Chae of Seoul National University, South Korea for his valuable discussions and suggestions.  R.A.M. thankfully acknowledge the support by the NITC/FRG-2019. }
		
	   \medskip{}
	   \noindent \textbf{Disclosure of Potential Conflicts of Interest.} The authors declare that they have no conflicts of interest.

	\end{article} 
\end{document}